\input harvmac

\input amssym
\input epsf

% FONTS
%\draftmode
% fraktur
\newfam\frakfam
\font\teneufm=eufm10
\font\seveneufm=eufm7
\font\fiveeufm=eufm5
\textfont\frakfam=\teneufm
\scriptfont\frakfam=\seveneufm
\scriptscriptfont\frakfam=\fiveeufm

% black board bold

\def\bb{
\font\tenmsb=msbm10
\font\sevenmsb=msbm7
\font\fivemsb=msbm5
\textfont1=\tenmsb
\scriptfont1=\sevenmsb
\scriptscriptfont1=\fivemsb
}

% double stroke math

\newfam\dsromfam
\font\tendsrom=dsrom10
\textfont\dsromfam=\tendsrom
\def\ds{\fam\dsromfam \tendsrom}

% bold math italics

\newfam\mbffam
\font\tenmbf=cmmib10
\font\sevenmbf=cmmib7
\font\fivembf=cmmib5
\textfont\mbffam=\tenmbf
\scriptfont\mbffam=\sevenmbf
\scriptscriptfont\mbffam=\fivembf

% bold math cal

\newfam\mbfcalfam
\font\tenmbfcal=cmbsy10
\font\sevenmbfcal=cmbsy7
\font\fivembfcal=cmbsy5
\textfont\mbfcalfam=\tenmbfcal
\scriptfont\mbfcalfam=\sevenmbfcal
\scriptscriptfont\mbfcalfam=\fivembfcal

% math script

\newfam\mscrfam
\font\tenmscr=rsfs10
\font\sevenmscr=rsfs7
\font\fivemscr=rsfs5
\textfont\mscrfam=\tenmscr
\scriptfont\mscrfam=\sevenmscr
\scriptscriptfont\mscrfam=\fivemscr

% MACROS

% bras, kets, ...

% tilde, hat, bar, ...

\def\hat{\widehat}

\def\bar{\overline}
\def\b{\bar}
\def\bsq#1{{{\b{#1}}^{\lower 2.5pt\hbox{$\scriptstyle 2$}}}}
\def\bexp#1#2{{{\b{#1}}^{\lower 2.5pt\hbox{$\scriptstyle #2$}}}}
\def\dotexp#1#2{{{#1}^{\lower 2.5pt\hbox{$\scriptstyle #2$}}}}

% basic math
\def\IL{\relax{\rm I\kern-.18em L}}
\def\IH{\relax{\rm I\kern-.18em H}}
\def\IR{\relax{\rm I\kern-.18em R}}
\def\IC{\relax{\rm I\kern-0.54 em C}}

\def\rt2{\sqrt{2}}
\def\half {{1 \over 2}}

\def\det{\mathop{\rm det}}

\def\Tr{\mathop{\rm Tr}}

% bold greek characters

\font\tenbifull=cmmib10
\font\tenbimed=cmmib7
\font\tenbismall=cmmib5
\textfont9=\tenbifull \scriptfont9=\tenbimed
\scriptscriptfont9=\tenbismall

\mathchardef\bbGamma="7000
\mathchardef\bbDelta="7001
\mathchardef\bbPhi="7002
\mathchardef\bbAlpha="7003
\mathchardef\bbXi="7004
\mathchardef\bbPi="7005
\mathchardef\bbSigma="7006
\mathchardef\bbUpsilon="7007
\mathchardef\bbTheta="7008
\mathchardef\bbPsi="7009
\mathchardef\bbOmega="700A
\mathchardef\bbalpha="710B
\mathchardef\bbbeta="710C
\mathchardef\bbgamma="710D
\mathchardef\bbdelta="710E
\mathchardef\bbepsilon="710F
\mathchardef\bbzeta="7110
\mathchardef\bbeta="7111
\mathchardef\bbtheta="7112
\mathchardef\bbiota="7113
\mathchardef\bbkappa="7114
\mathchardef\bblambda="7115
\mathchardef\bbmu="7116
\mathchardef\bbnu="7117
\mathchardef\bbxi="7118
\mathchardef\bbpi="7119
\mathchardef\bbrho="711A
\mathchardef\bbsigma="711B
\mathchardef\bbtau="711C
\mathchardef\bbupsilon="711D
\mathchardef\bbphi="711E
\mathchardef\bbchi="711F
\mathchardef\bbpsi="7120
\mathchardef\bbomega="7121
\mathchardef\bbvarepsilon="7122
\mathchardef\bbvartheta="7123
\mathchardef\bbvarpi="7124
\mathchardef\bbvarrho="7125
\mathchardef\bbvarsigma="7126
\mathchardef\bbvarphi="7127

\def\IL{\relax{\rm I\kern-.18em L}}
\def\IH{\relax{\rm I\kern-.18em H}}
\def\IR{\relax{\rm I\kern-.18em R}}
\def\IC{\relax{\rm I\kern-0.54 em C}}

% dotted spinor indices

% bared indices

% bared spinors

% capital cal letters

\def\CA{{\cal A}}

\def\CC{{\cal C}}

\def\CH{{\cal H}}

\def\CN{{\cal N}}
\def\CO{{\cal O}}
\def\CP{{\cal P}}

\def\CT{{\cal T}}

% double stroke symbols: unit matrix, reals, complex, quaternions, integers, natural numbers

\def\1{{\ds 1}}
\def\R{\hbox{$\bb R$}}
\def\C{\hbox{$\bb C$}}

\def\Z{\hbox{$\bb Z$}}

\def\P{\hbox{$\bb P$}}
\def\Sn{\hbox{$\bb S$}}

% miscellaneous objects

\noblackbox

\def\unit{\relax{\rm 1\kern-.26em I}}
\def\nada{\relax{\rm 0\kern-.30em l}}

\def\det{{\rm det}}
\def\CP{{\cal P}}
%% MACROS
\noblackbox
\def\IL{\relax{\rm I\kern-.18em L}}
\def\IH{\relax{\rm I\kern-.18em H}}
\def\IR{\relax{\rm I\kern-.18em R}}
\def\IC{\relax\hbox{$\inbar\kern-.3em{\rm C}$}}
\def\IZ{\relax\ifmmode\mathchoice
{\hbox{\cmss Z\kern-.4em Z}}{\hbox{\cmss Z\kern-.4em Z}} {\lower.9pt\hbox{\cmsss Z\kern-.4em Z}}
{\lower1.2pt\hbox{\cmsss Z\kern-.4em Z}}\else{\cmss Z\kern-.4em Z}\fi}

\def\CN {{\cal N}}

\def\partialslash{\not{\hbox{\kern-2pt $\partial$}}}
\def\CP {{\cal P }}

\def\CO {{\cal O}}

\def\CH {{\cal H}}
\def\CC {{\cal C}}

\def\CA{{\cal A}}

%% MORE MACROS

\def\CN {{\cal N}}

\def\CO {{\cal O}}

\def\CP {{\cal P }}

\def\det{{\rm det}}
\def\Tr{{\rm Tr}}

\font\manual=manfnt \def\dbend{\lower3.5pt\hbox{\manual\char127}}

\def\IZ{\relax\ifmmode\mathchoice
{\hbox{\cmss Z\kern-.4em Z}}{\hbox{\cmss Z\kern-.4em Z}} {\lower.9pt\hbox{\cmsss Z\kern-.4em Z}}
{\lower1.2pt\hbox{\cmsss Z\kern-.4em Z}}\else{\cmss Z\kern-.4em Z}\fi}
\def\half {{1\over 2}}
\def\sdtimes{\mathbin{\hbox{\hskip2pt\vrule height 4.1pt depth -.3pt
width .25pt \hskip-2pt$\times$}}  }

\def\bar{\overline}

\def\CH{{\cal H}}

\def\rt2{\sqrt{2}}
\def\irt2{{1\over\sqrt{2}}}

\def\hat{\widehat}
%  \slashchar puts a slash through a character to represent contraction
%  with Dirac matrices. Use \not instead for negation of relations, and use
%  \hbar for hbar.
\def\slashchar#1{\setbox0=\hbox{$#1$}           % set a box for #1
   \dimen0=\wd0                                 % and get its size
   \setbox1=\hbox{/} \dimen1=\wd1               % get size of /
   \ifdim\dimen0>\dimen1                        % #1 is bigger
      \rlap{\hbox to \dimen0{\hfil/\hfil}}      % so center / in box
      #1                                        % and print #1
   \else                                        % / is bigger
      \rlap{\hbox to \dimen1{\hfil$#1$\hfil}}   % so center #1
      /                                         % and print /
   \fi}

\def\foursqr#1#2{{\vcenter{\vbox{
    \hrule height.#2pt
    \hbox{\vrule width.#2pt height#1pt \kern#1pt
    \vrule width.#2pt}
    \hrule height.#2pt
    \hrule height.#2pt
    \hbox{\vrule width.#2pt height#1pt \kern#1pt
    \vrule width.#2pt}
    \hrule height.#2pt
        \hrule height.#2pt
    \hbox{\vrule width.#2pt height#1pt \kern#1pt
    \vrule width.#2pt}
    \hrule height.#2pt
        \hrule height.#2pt
    \hbox{\vrule width.#2pt height#1pt \kern#1pt
    \vrule width.#2pt}
    \hrule height.#2pt}}}}
\def\psqr#1#2{{\vcenter{\vbox{\hrule height.#2pt
    \hbox{\vrule width.#2pt height#1pt \kern#1pt
    \vrule width.#2pt}
    \hrule height.#2pt \hrule height.#2pt
    \hbox{\vrule width.#2pt height#1pt \kern#1pt
    \vrule width.#2pt}
    \hrule height.#2pt}}}}
\def\sqr#1#2{{\vcenter{\vbox{\hrule height.#2pt
    \hbox{\vrule width.#2pt height#1pt \kern#1pt
    \vrule width.#2pt}
    \hrule height.#2pt}}}}

\def\figin{\epsfcheck\figin}\def\figins{\epsfcheck\figins}
\def\epsfcheck{\ifx\epsfbox\UnDeFiNeD
\message{(NO epsf.tex, FIGURES WILL BE IGNORED)}
\gdef\figin##1{\vskip2in}\gdef\figins##1{\hskip.5in}% blank space instead
\else\message{(FIGURES WILL BE INCLUDED)}%
\gdef\figin##1{##1}\gdef\figins##1{##1}\fi}
\def\DefWarn#1{}
\def\figinsert{\goodbreak\midinsert}
\def\ifig#1#2#3{\DefWarn#1\xdef#1{fig.~\the\figno}
\writedef{#1\leftbracket fig.\noexpand~\the\figno}%
\figinsert\figin{\centerline{#3}}\medskip\centerline{\vbox{\baselineskip12pt \advance\hsize by
-1truein\noindent\footnotefont{\bf Fig.~\the\figno:\ } \it#2}}
\bigskip\endinsert\global\advance\figno by1}

%\VenezianoYB
\lref\VenezianoYB{
  G.~Veneziano,
  ``Construction of a crossing - symmetric, Regge behaved amplitude for linearly rising trajectories,''
Nuovo Cim.\ A {\bf 57}, 190 (1968).
}

%\GrossBR
\lref\GrossBR{
  D.~J.~Gross, R.~D.~Pisarski and L.~G.~Yaffe,
  ``QCD and Instantons at Finite Temperature,''
Rev.\ Mod.\ Phys.\  {\bf 53}, 43 (1981).
%%CITATION = PRINT-80-0538 (PRINCETON)%%
}

%\SvetitskyGS
\lref\SvetitskyGS{
  B.~Svetitsky and L.~G.~Yaffe,
  ``Critical Behavior at Finite Temperature Confinement Transitions,''
Nucl.\ Phys.\ B {\bf 210}, 423 (1982).
%%CITATION = CLNS-82/530%%
}

%\SvetitskyYE
\lref\SvetitskyYE{
  B.~Svetitsky,
  ``Symmetry Aspects of Finite Temperature Confinement Transitions,''
Phys.\ Rept.\  {\bf 132}, 1 (1986).
%%CITATION = MIT-CTP-1245%%
}

%\SiversIG
\lref\SiversIG{
  D.~Sivers and J.~Yellin,
  ``Review of recent work on narrow resonance models,''
Rev.\ Mod.\ Phys.\  {\bf 43}, 125 (1971).
}

%\WittenEY
\lref\WittenEY{
  E.~Witten,
  ``Dyons of Charge e theta/2 pi,''
Phys.\ Lett.\ B {\bf 86}, 283 (1979).
%%CITATION = CERN-TH-2724%%
}

%\GreenSG
\lref\GreenSG{
  M.~B.~Green and J.~H.~Schwarz,
  ``Anomaly Cancellation in Supersymmetric D=10 Gauge Theory and Superstring Theory,''
Phys.\ Lett.\  {\bf 149B}, 117 (1984).
%%CITATION = CALT-68-1182%%
}

%\ColemanUZ
\lref\ColemanUZ{
  S.~R.~Coleman,
  ``More About the Massive Schwinger Model,''
Annals Phys.\  {\bf 101}, 239 (1976).
%%CITATION = Print-76-0357 (HARVARD)%%
}

%\BaluniRF
\lref\BaluniRF{
  V.~Baluni,
  ``CP Violating Effects in QCD,''
Phys.\ Rev.\ D {\bf 19}, 2227 (1979).
%%CITATION = MIT-CTP-726%%
}

%\DashenET
\lref\DashenET{
  R.~F.~Dashen,
  ``Some features of chiral symmetry breaking,''
Phys.\ Rev.\ D {\bf 3}, 1879 (1971).
}

%\WittenBC
\lref\WittenBC{
  E.~Witten,
  ``Instantons, the Quark Model, and the 1/n Expansion,''
Nucl.\ Phys.\ B {\bf 149}, 285 (1979).
%%CITATION = HUTP-78/A042%%
}

%\WittenVV
\lref\WittenVV{
  E.~Witten,
  ``Current Algebra Theorems for the U(1) Goldstone Boson,''
Nucl.\ Phys.\ B {\bf 156}, 269 (1979).
%%CITATION = HUTP-79/A014%%
}

%\WittenSP
\lref\WittenSP{
  E.~Witten,
  ``Large N Chiral Dynamics,''
Annals Phys.\  {\bf 128}, 363 (1980).
%%CITATION = HUTP-80/A005%%
}

%\DAddaVBW
\lref\DAddaVBW{
  A.~D'Adda, M.~Luscher and P.~Di Vecchia,
  ``A 1/n Expandable Series of Nonlinear Sigma Models with Instantons,''
Nucl.\ Phys.\ B {\bf 146}, 63 (1978).
%%CITATION = Print-78-0885 (BOHR INST.)%%
}

%\AcharyaDZ
\lref\AcharyaDZ{
  B.~S.~Acharya and C.~Vafa,
  ``On domain walls of N=1 supersymmetric Yang-Mills in four-dimensions,''
[hep-th/0103011].
%%CITATION = hep-th/0103011%%
}

%\AffleckCH
\lref\AffleckCH{
  I.~Affleck and F.~D.~M.~Haldane,
  ``Critical Theory of Quantum Spin Chains,''
Phys.\ Rev.\ B {\bf 36}, 5291 (1987).
%%CITATION = PUPT-1052%%
}

%\BilloJDA
\lref\BilloJDA{
  M.~Billó, M.~Caselle, D.~Gaiotto, F.~Gliozzi, M.~Meineri and R.~Pellegrini,
  ``Line defects in the 3d Ising model,''
JHEP {\bf 1307}, 055 (2013).
[arXiv:1304.4110 [hep-th]].
%%CITATION = arXiv:1304.4110%%
}

%\WittenABA
\lref\WittenABA{
  E.~Witten,
  ``Fermion Path Integrals And Topological Phases,''
Rev.\ Mod.\ Phys.\  {\bf 88}, no. 3, 035001 (2016).
[arXiv:1508.04715 [cond-mat.mes-hall]].
%%CITATION = arXiv:1508.04715%%
}

%\GaiottoNVA
\lref\GaiottoNVA{
  D.~Gaiotto, D.~Mazac and M.~F.~Paulos,
  min
  ``Bootstrapping the 3d Ising twist defect,''
JHEP {\bf 1403}, 100 (2014).
[arXiv:1310.5078 [hep-th]].
%%CITATION = arXiv:1310.5078%%
}

%\BrowerEA
\lref\BrowerEA{
  R.~C.~Brower, J.~Polchinski, M.~J.~Strassler and C.~I.~Tan,
  ``The Pomeron and gauge/string duality,''
JHEP {\bf 0712}, 005 (2007).
[hep-th/0603115].
%%CITATION = hep-th/0603115%%
}

\lref\inprogress{
  In progress.
%%CITATION = hep-th/0603115%%
}

\lref\inprogressi{
  In progress.
%%CITATION = hep-th/0603115%%
}

\lref\mandelstam{
S.~Mandelstam, ``Dual-resonance models." Physics Reports 13.6 (1974): 259-353.
}

%\FreundHW
\lref\FreundHW{
  P.~G.~O.~Freund,
  ``Finite energy sum rules and bootstraps,''
Phys.\ Rev.\ Lett.\  {\bf 20}, 235 (1968).
}

%\MeyerJC
\lref\MeyerJC{
  H.~B.~Meyer and M.~J.~Teper,
  ``Glueball Regge trajectories and the pomeron: A Lattice study,''
Phys.\ Lett.\ B {\bf 605}, 344 (2005).
[hep-ph/0409183].
%%CITATION = hep-ph/0409183%%
}

%\CoonYW
\lref\CoonYW{
  D.~D.~Coon,
  ``Uniqueness of the veneziano representation,''
Phys.\ Lett.\ B {\bf 29}, 669 (1969).
}

%\FairlieAD
\lref\FairlieAD{
  D.~B.~Fairlie and J.~Nuyts,
  ``A fresh look at generalized Veneziano amplitudes,''
Nucl.\ Phys.\ B {\bf 433}, 26 (1995).
[hep-th/9406043].
%%CITATION = hep-th/9406043%%
}

%\C
\lref\CWpot{
  S.~R.~Coleman and E.~J.~Weinberg,
  ``Radiative Corrections as the Origin of Spontaneous Symmetry Breaking,''
Phys.\ Rev.\ D {\bf 7}, 1888 (1973).
}

%\RedlichDV
\lref\RedlichDV{
  A.~N.~Redlich,
  ``Parity Violation and Gauge Noninvariance of the Effective Gauge Field Action in Three-Dimensions,''
Phys.\ Rev.\ D {\bf 29}, 2366 (1984).
%%CITATION = MIT-CTP-1128%%
}

%\RedlichKN
\lref\RedlichKN{
  A.~N.~Redlich,
  ``Gauge Noninvariance and Parity Violation of Three-Dimensional Fermions,''
Phys.\ Rev.\ Lett.\  {\bf 52}, 18 (1984).
%%CITATION = MIT-CTP-1107%%
}

%\PonomarevJQK
\lref\PonomarevJQK{
  D.~Ponomarev and A.~A.~Tseytlin,
  %``On quantum corrections in higher-spin theory in flat space,''
[arXiv:1603.06273 [hep-th]].
%%CITATION = Imperial-TP-DP-2016-01%%
}

\lref\StromingerTalk{
  A.~Strominger, Talk at Strings 2014, Princeton.
%%CITATION = Imperial-TP-DP-2016-01%%
}

%\PoppitzNZ
\lref\PoppitzNZ{
  E.~Poppitz, T.~Schäfer and M.~Ünsal,
  ``Universal mechanism of (semi-classical) deconfinement and theta-dependence for all simple groups,''
JHEP {\bf 1303}, 087 (2013).
[arXiv:1212.1238 [hep-th]].
%%CITATION = arXiv:1212.1238%%
}

%\UnsalZJ
\lref\UnsalZJ{
  M.~Unsal,
  ``Theta dependence, sign problems and topological interference,''
Phys.\ Rev.\ D {\bf 86}, 105012 (2012).
[arXiv:1201.6426 [hep-th]].
%%CITATION = SLAC-PUB-16017%%
}

%\CostaMG
\lref\CostaMG{
  M.~S.~Costa, J.~Penedones, D.~Poland and S.~Rychkov,
  %``Spinning Conformal Correlators,''
JHEP {\bf 1111}, 071 (2011).
[arXiv:1107.3554 [hep-th]].
%%CITATION = LPTENS-11-22%%
}

%\CamanhoAPA
\lref\CamanhoAPA{
  X.~O.~Camanho, J.~D.~Edelstein, J.~Maldacena and A.~Zhiboedov,
  %``Causality Constraints on Corrections to the Graviton Three-Point Coupling,''
JHEP {\bf 1602}, 020 (2016).
[arXiv:1407.5597 [hep-th]].
%%CITATION = arXiv:1407.5597%%
}

%\LandsteinerCP
\lref\LandsteinerCP{
  K.~Landsteiner, E.~Megias and F.~Pena-Benitez,
  ``Gravitational Anomaly and Transport,''
Phys.\ Rev.\ Lett.\  {\bf 107}, 021601 (2011).
[arXiv:1103.5006 [hep-ph]].
%%CITATION = arXiv:1103.5006%%
}

%\BanerjeeIZ
\lref\BanerjeeIZ{
  N.~Banerjee, J.~Bhattacharya, S.~Bhattacharyya, S.~Jain, S.~Minwalla and T.~Sharma,
  ``Constraints on Fluid Dynamics from Equilibrium Partition Functions,''
JHEP {\bf 1209}, 046 (2012).
[arXiv:1203.3544 [hep-th]].
%%CITATION = TFR-TH-12-05%%
}

%\JensenKJ
\lref\JensenKJ{
  K.~Jensen, R.~Loganayagam and A.~Yarom,
  ``Thermodynamics, gravitational anomalies and cones,''
JHEP {\bf 1302}, 088 (2013).
[arXiv:1207.5824 [hep-th]].
%%CITATION = arXiv:1207.5824%%
}

%\BonettiELA
\lref\BonettiELA{
  F.~Bonetti, T.~W.~Grimm and S.~Hohenegger,
  ``One-loop Chern-Simons terms in five dimensions,''
JHEP {\bf 1307}, 043 (2013).
[arXiv:1302.2918 [hep-th]].
%%CITATION = arXiv:1302.2918%%
}

%\DiPietroBCA
\lref\DiPietroBCA{
  L.~Di Pietro and Z.~Komargodski,
  ``Cardy formulae for SUSY theories in $d =$ 4 and $d =$ 6,''
JHEP {\bf 1412}, 031 (2014).
[arXiv:1407.6061 [hep-th]].
%%CITATION = arXiv:1407.6061%%
}

%\ArdehaliHYA
\lref\ArdehaliHYA{
  A.~Arabi Ardehali, J.~T.~Liu and P.~Szepietowski,
  ``High-Temperature Expansion of Supersymmetric Partition Functions,''
JHEP {\bf 1507}, 113 (2015).
[arXiv:1502.07737 [hep-th]].
%%CITATION = MCTP-15-06%%
}

%\FeiOHA
\lref\FeiOHA{
  L.~Fei, S.~Giombi, I.~R.~Klebanov and G.~Tarnopolsky,
  ``Generalized $F$-Theorem and the $\epsilon$ Expansion,''
JHEP {\bf 1512}, 155 (2015).
[arXiv:1507.01960 [hep-th]].
%%CITATION = PUPT-2481%%
}

%\AcharyaDZ
\lref\AcharyaDZ{
  B.~S.~Acharya and C.~Vafa,
  ``On domain walls of N=1 supersymmetric Yang-Mills in four-dimensions,''
[hep-th/0103011].
%%CITATION = hep-th/0103011%%
}

%\GiombiXXA
\lref\GiombiXXA{
  S.~Giombi and I.~R.~Klebanov,
  ``Interpolating between $a$ and $F$,''
JHEP {\bf 1503}, 117 (2015).
[arXiv:1409.1937 [hep-th]].
%%CITATION = PUPT-2472%%
}

%\KapustinGUA
\lref\KapustinGUA{
  A.~Kapustin and N.~Seiberg,
  ``Coupling a QFT to a TQFT and Duality,''
JHEP {\bf 1404}, 001 (2014).
[arXiv:1401.0740 [hep-th]].
%%CITATION = arXiv:1401.0740%%
}

%\GaiottoKFA
\lref\GaiottoKFA{
  D.~Gaiotto, A.~Kapustin, N.~Seiberg and B.~Willett,
  ``Generalized Global Symmetries,''
JHEP {\bf 1502}, 172 (2015).
[arXiv:1412.5148 [hep-th]].
%%CITATION = arXiv:1412.5148%%
}

%\DijkgraafPZ
\lref\DijkgraafPZ{
  R.~Dijkgraaf and E.~Witten,
  ``Topological Gauge Theories and Group Cohomology,''
Commun.\ Math.\ Phys.\  {\bf 129}, 393 (1990).
%%CITATION = THU-89-9%%
}

%\KadanoffKZ
\lref\KadanoffKZ{
  L.~P.~Kadanoff and H.~Ceva,
  ``Determination of an opeator algebra for the two-dimensional Ising model,''
Phys.\ Rev.\ B {\bf 3}, 3918 (1971).
}

%\GrossKZA
\lref\GrossKZA{
  D.~J.~Gross and P.~F.~Mende,
  %``The High-Energy Behavior of String Scattering Amplitudes,''
Phys.\ Lett.\ B {\bf 197}, 129 (1987).
%%CITATION = PUPT-1062%%
}

%\KarlinerHD
\lref\KarlinerHD{
  M.~Karliner, I.~R.~Klebanov and L.~Susskind,
  %``Size and Shape of Strings,''
Int.\ J.\ Mod.\ Phys.\ A {\bf 3}, 1981 (1988).
%%CITATION = SLAC-PUB-4531%%
}

%\ShankarEE
\lref\Shankar{
  R.~Shankar and N.~Read,
  ``The $\theta = \pi$ Nonlinear $\sigma$ Model Is Massless,''
Nucl.\ Phys.\ B {\bf 336}, 457 (1990).
%%CITATION = NSF-ITP-89-98%%
}

%\CachazoZK
\lref\CachazoZK{
  F.~Cachazo, N.~Seiberg and E.~Witten,
  ``Phases of N=1 supersymmetric gauge theories and matrices,''
JHEP {\bf 0302}, 042 (2003).
[hep-th/0301006].
%%CITATION = hep-th/0301006%%
}

%\WeissRJ
\lref\WeissRJ{
  N.~Weiss,
  ``The Effective Potential for the Order Parameter of Gauge Theories at Finite Temperature,''
Phys.\ Rev.\ D {\bf 24}, 475 (1981).
%%CITATION = UBC-81%%
}

%\DashenET
\lref\DashenET{
  R.~F.~Dashen,
  ``Some features of chiral symmetry breaking,''
Phys.\ Rev.\ D {\bf 3}, 1879 (1971).
}

\lref\Hitoshi{
http://hitoshi.berkeley.edu/221b/scattering3.pdf
}

%\GreensiteZZ
\lref\GreensiteZZ{
  J.~Greensite,
  ``An introduction to the confinement problem,''
Lect.\ Notes Phys.\  {\bf 821}, 1 (2011).
}

%\SeibergRS
\lref\SeibergRS{
  N.~Seiberg and E.~Witten,
  ``Electric - magnetic duality, monopole condensation, and confinement in N=2 supersymmetric Yang-Mills theory,''
Nucl.\ Phys.\ B {\bf 426}, 19 (1994), Erratum: [Nucl.\ Phys.\ B {\bf 430}, 485 (1994)].
[hep-th/9407087].
%%CITATION = hep-th/9407087%%
}

%\ColemanUZ
\lref\ColemanUZ{
  S.~R.~Coleman,
  ``More About the Massive Schwinger Model,''
Annals Phys.\  {\bf 101}, 239 (1976).
%%CITATION = Print-76-0357 (HARVARD)%%
}
%\IntriligatorAU
\lref\IntriligatorAU{
  K.~A.~Intriligator and N.~Seiberg,
  ``Lectures on supersymmetric gauge theories and electric-magnetic duality,''
Nucl.\ Phys.\ Proc.\ Suppl.\  {\bf 45BC}, 1 (1996), [Subnucl.\ Ser.\  {\bf 34}, 237 (1997)].
[hep-th/9509066].
%%CITATION = hep-th/9509066%%
}

%\SeibergAJ
\lref\SeibergAJ{
  N.~Seiberg and E.~Witten,
  ``Monopoles, duality and chiral symmetry breaking in N=2 supersymmetric QCD,''
Nucl.\ Phys.\ B {\bf 431}, 484 (1994).
[hep-th/9408099].
%%CITATION = hep-th/9408099%%
}

\lref\Deepak{
D.~Naidu, ``Categorical Morita equivalence for group-theoretical categories, " Communications in Algebra 35.11 (2007): 3544-3565.
APA	
}
%\AharonyDHA
\lref\AharonyDHA{
  O.~Aharony, S.~S.~Razamat, N.~Seiberg and B.~Willett,
  ``3d dualities from 4d dualities,''
JHEP {\bf 1307}, 149 (2013).
[arXiv:1305.3924 [hep-th]].
%%CITATION = WIS-04-13-APR-DPPA%%
}

%\ArgyresJJ
\lref\ArgyresJJ{
  P.~C.~Argyres and M.~R.~Douglas,
  ``New phenomena in SU(3) supersymmetric gauge theory,''
Nucl.\ Phys.\ B {\bf 448}, 93 (1995).
[hep-th/9505062].
%%CITATION = hep-th/9505062%%
}

%\SusskindAA
\lref\SusskindAA{
  L.~Susskind,
  %``Strings, black holes and Lorentz contraction,''
Phys.\ Rev.\ D {\bf 49}, 6606 (1994).
[hep-th/9308139].
%%CITATION = hep-th/9308139%%
}

%\DubovskySH
\lref\DubovskySH{
  S.~Dubovsky, R.~Flauger and V.~Gorbenko,
  ``Effective String Theory Revisited,''
JHEP {\bf 1209}, 044 (2012).
[arXiv:1203.1054 [hep-th]].
%%CITATION = arXiv:1203.1054%%
}

%\AharonyIPA
\lref\AharonyIPA{
  O.~Aharony and Z.~Komargodski,
  ``The Effective Theory of Long Strings,''
JHEP {\bf 1305}, 118 (2013).
[arXiv:1302.6257 [hep-th]].
%%CITATION = WIS-01-13-FEB-DPPA%%
}

%\HellermanCBA
\lref\HellermanCBA{
  S.~Hellerman, S.~Maeda, J.~Maltz and I.~Swanson,
  ``Effective String Theory Simplified,''
JHEP {\bf 1409}, 183 (2014).
[arXiv:1405.6197 [hep-th]].
%%CITATION = IPMU-14-0125%%
}

%\AthenodorouCS
\lref\AthenodorouCS{
  A.~Athenodorou, B.~Bringoltz and M.~Teper,
  ``Closed flux tubes and their string description in D=3+1 SU(N) gauge theories,''
JHEP {\bf 1102}, 030 (2011).
[arXiv:1007.4720 [hep-lat]].
%%CITATION = arXiv:1007.4720%%
}

%\AharonyHDA
\lref\AharonyHDA{
  O.~Aharony, N.~Seiberg and Y.~Tachikawa,
  ``Reading between the lines of four-dimensional gauge theories,''
JHEP {\bf 1308}, 115 (2013).
[arXiv:1305.0318 [hep-th]].
%%CITATION = UT-13-15%%
}

%\CallanSA
\lref\CallanSA{
  C.~G.~Callan, Jr. and J.~A.~Harvey,
  ``Anomalies and Fermion Zero Modes on Strings and Domain Walls,''
Nucl.\ Phys.\ B {\bf 250}, 427 (1985).
%%CITATION = Print-84-0860 (PRINCETON)%%
}

%\tHooftXSS
\lref\tHooftXSS{
  G.~'t Hooft et al.,
    ``Recent Developments in Gauge Theories. Proceedings, Nato Advanced Study Institute, Cargese, France, August 26 - September 8, 1979,''
NATO Sci.\ Ser.\ B {\bf 59}, pp.1 (1980).
}

%\AppelquistVG
\lref\AppelquistVG{
  T.~Appelquist and R.~D.~Pisarski,
  ``High-Temperature Yang-Mills Theories and Three-Dimensional Quantum Chromodynamics,''
Phys.\ Rev.\ D {\bf 23}, 2305 (1981).
%%CITATION = Print-81-0020 (YALE)%%
}

%\HarveyIT
\lref\HarveyIT{
  J.~A.~Harvey,
  ``TASI 2003 lectures on anomalies,''
[hep-th/0509097].
%%CITATION = hep-th/0509097%%
}

%\KonishiIZ
\lref\KonishiIZ{
  K.~Konishi,
  ``Confinement, supersymmetry breaking and theta parameter dependence in the Seiberg-Witten model,''
Phys.\ Lett.\ B {\bf 392}, 101 (1997).
[hep-th/9609021].
%%CITATION = hep-th/9609021%%
}

%\DineSGQ
\lref\DineSGQ{
  M.~Dine, P.~Draper, L.~Stephenson-Haskins and D.~Xu,
  ``$\theta$ and the $\eta^\prime$ in Large $N$ Supersymmetric QCD,''
[arXiv:1612.05770 [hep-th]].
%%CITATION = arXiv:1612.05770%%
}

%\'tHooftUJ
\lref\tHooftUJ{
  G.~'t Hooft,
  ``A Property of Electric and Magnetic Flux in Nonabelian Gauge Theories,''
Nucl.\ Phys.\ B {\bf 153}, 141 (1979).
%%CITATION = PRINT-79-0117 (UTRECHT)%%
}

\lref\Brower{
R.~C.~Brower and J.~Harte.
%``Kinematic Constraints for Infinitely Rising Regge Trajectories,"
Physical Review 164.5 (1967): 1841.
}

%\ArgyresXN
\lref\ArgyresXN{
  P.~C.~Argyres, M.~R.~Plesser, N.~Seiberg and E.~Witten,
  ``New N=2 superconformal field theories in four-dimensions,''
Nucl.\ Phys.\ B {\bf 461}, 71 (1996).
[hep-th/9511154].
%%CITATION = hep-th/9511154%%
}

%\JakubskyKI
\lref\JakubskyKI{
  V.~Jakubsky, L.~M.~Nieto and M.~S.~Plyushchay,
  ``The origin of the hidden supersymmetry,''
Phys.\ Lett.\ B {\bf 692}, 51 (2010).
[arXiv:1004.5489 [hep-th]].
%%CITATION = arXiv:1004.5489%%
}

%\CorreaJE
\lref\CorreaJE{
  F.~Correa and M.~S.~Plyushchay,
  ``Hidden supersymmetry in quantum bosonic systems,''
Annals Phys.\  {\bf 322}, 2493 (2007).
[hep-th/0605104].
%%CITATION = hep-th/0605104%%
}

%\HenningsonHP
\lref\HenningsonHP{
  M.~Henningson,
  ``Wilson-'t Hooft operators and the theta angle,''
JHEP {\bf 0605}, 065 (2006).
[hep-th/0603188].
%%CITATION = hep-th/0603188%%
}

%\CreutzXU
\lref\CreutzXU{
  M.~Creutz,
  ``Spontaneous violation of CP symmetry in the strong interactions,''
Phys.\ Rev.\ Lett.\  {\bf 92}, 201601 (2004).
[hep-lat/0312018].
%%CITATION = hep-lat/0312018%%
}

%\AffleckTJ
\lref\AffleckTJ{
  I.~Affleck,
  ``Nonlinear sigma model at Theta = pi: Euclidean lattice formulation and solid-on-solid models,''
Phys.\ Rev.\ Lett.\  {\bf 66}, 2429 (1991).
%%CITATION = UBCTP-91-005%%
}

%\ChenPG
\lref\ChenPG{
  X.~Chen, Z.~C.~Gu, Z.~X.~Liu and X.~G.~Wen,
  ``Symmetry protected topological orders and the group cohomology of their symmetry group,''
Phys.\ Rev.\ B {\bf 87}, no. 15, 155114 (2013).
[arXiv:1106.4772 [cond-mat.str-el]].
%%CITATION = arXiv:1106.4772%%
}

%\PappadopuloJK
\lref\PappadopuloJK{
  D.~Pappadopulo, S.~Rychkov, J.~Espin and R.~Rattazzi,
  %``OPE Convergence in Conformal Field Theory,''
Phys.\ Rev.\ D {\bf 86}, 105043 (2012).
[arXiv:1208.6449 [hep-th]].
%%CITATION = LPTENS-12-31%%
}

\lref\LegRed{
Askey, Richard. "Orthogonal expansions with positive coefficients." Proceedings of the American Mathematical Society 16.6 (1965): 1191-1194.
}

%\ClossetVP
\lref\ClossetVP{
  C.~Closset, T.~T.~Dumitrescu, G.~Festuccia, Z.~Komargodski and N.~Seiberg,
  ``Comments on Chern-Simons Contact Terms in Three Dimensions,''
JHEP {\bf 1209}, 091 (2012).
[arXiv:1206.5218 [hep-th]].
%%CITATION = PUTP-2417%%
}

%\DieriglXTA
\lref\DieriglXTA{
  M.~Dierigl and A.~Pritzel,
  ``Topological Model for Domain Walls in (Super-)Yang-Mills Theories,''
Phys.\ Rev.\ D {\bf 90}, no. 10, 105008 (2014).
[arXiv:1405.4291 [hep-th]].
%%CITATION = arXiv:1405.4291%%
}

%\WittenDF
\lref\WittenDF{
  E.~Witten,
  ``Constraints on Supersymmetry Breaking,''
Nucl.\ Phys.\ B {\bf 202}, 253 (1982).
%%CITATION = PRINT-82-0163 (PRINCETON)%%
}

%\SeibergRSG
\lref\SeibergRSG{
  N.~Seiberg and E.~Witten,
  ``Gapped Boundary Phases of Topological Insulators via Weak Coupling,''
PTEP {\bf 2016}, no. 12, 12C101 (2016).
[arXiv:1602.04251 [cond-mat.str-el]].
%%CITATION = arXiv:1602.04251%%
}

%\BeniniDUS
\lref\BeniniDUS{
  F.~Benini, P.~S.~Hsin and N.~Seiberg,
  ``Comments on Global Symmetries, Anomalies, and Duality in (2+1)d,''
[arXiv:1702.07035 [cond-mat.str-el]].
%%CITATION = arXiv:1702.07035%%
}

%\AmatiWQ
\lref\AmatiWQ{
  D.~Amati, M.~Ciafaloni and G.~Veneziano,
  %``Superstring Collisions at Planckian Energies,''
Phys.\ Lett.\ B {\bf 197}, 81 (1987).
%%CITATION = CERN-TH-4782/87%%
}

%\HsinBLU
\lref\HsinBLU{
  P.~S.~Hsin and N.~Seiberg,
  ``Level/rank Duality and Chern-Simons-Matter Theories,''
JHEP {\bf 1609}, 095 (2016).
[arXiv:1607.07457 [hep-th]].
%%CITATION = arXiv:1607.07457%%
}

%\KapustinLWA
\lref\KapustinLWA{
  A.~Kapustin and R.~Thorngren,
  ``Anomalies of discrete symmetries in three dimensions and group cohomology,''
Phys.\ Rev.\ Lett.\  {\bf 112}, no. 23, 231602 (2014).
[arXiv:1403.0617 [hep-th]].
%%CITATION = arXiv:1403.0617%%
}

%\ElitzurXJ
\lref\ElitzurXJ{
  S.~Elitzur, Y.~Frishman, E.~Rabinovici and A.~Schwimmer,
  ``Origins of Global Anomalies in Quantum Mechanics,''
Nucl.\ Phys.\ B {\bf 273}, 93 (1986).
%%CITATION = WIS-85/48-Ph%%
}

%\KapustinZVA
\lref\KapustinZVA{
  A.~Kapustin and R.~Thorngren,
  ``Anomalies of discrete symmetries in various dimensions and group cohomology,''
[arXiv:1404.3230 [hep-th]].
%%CITATION = arXiv:1404.3230%%
}

\lref\Hatcher{
A.~Hatcher, ``Algebraic topology,'' Cambridge University Press, 2002.
}

%\ZamolodchikovZR
\lref\ZamolodchikovZR{
  A.~B.~Zamolodchikov and A.~B.~Zamolodchikov,
  ``Massless factorized scattering and sigma models with topological terms,''
Nucl.\ Phys.\ B {\bf 379}, 602 (1992).
}

%\draftmode

\vskip-20pt
\Title{
} {\vbox{\centerline{Theta, Time Reversal, and Temperature  }
}}
\vskip-20pt
\centerline{Davide Gaiotto,\foot{ Perimeter Institute for Theoretical Physics,
Waterloo, Ontario, N2L 2Y5, Canada.} Anton Kapustin,\foot{Walter Burke Institute for Theoretical Physics, California Institute of Technology, Pasadena, CA 91125.} Zohar Komargodski,\foot{Weizmann Institute of Science, Rehovot 76100, Israel.}  and  Nathan Seiberg\foot{School of Natural Sciences, Institute for Advanced Study, Princeton, NJ 08540, USA.}}
\vskip15pt
\vskip5pt

\noindent
$SU(N)$ gauge theory is time reversal invariant at $\theta=0$ and $\theta=\pi$. We show that at $\theta=\pi$ there is a discrete 't Hooft anomaly involving time reversal and the center symmetry.  This anomaly leads to constraints on the vacua of the theory. It follows that at $\theta=\pi$ the vacuum cannot be a trivial non-degenerate gapped state. (By contrast, the vacuum at $\theta=0$ is gapped, non-degenerate, and trivial.)  Due to the anomaly, the theory admits nontrivial domain walls supporting lower-dimensional theories.
Depending on the nature of the vacuum at $\theta=\pi$, several phase diagrams are possible.
Assuming area law for space-like loops, one arrives at an inequality involving the temperatures at which CP and the center symmetry are restored. We also analyze alternative scenarios for $SU(2)$ gauge theory. The underlying symmetry at $\theta=\pi$ is the dihedral group of 8 elements. If deconfined loops are allowed, one can have two $O(2)$-symmetric fixed points.  It may also be that the four-dimensional theory around $\theta=\pi$ is gapless, e.g.\ a Coulomb phase could match the underlying anomalies.

\vskip10pt

\Date{February 2017}

\newsec{Introduction}

One of the central tools for analyzing strongly coupled systems is 't Hooft's anomaly matching~\tHooftXSS. Given a theory with symmetry $G$, we may try to couple $G$ to classical background gauge fields. It is sometimes impossible to do that in spite of the fact that $G$ is a true symmetry of the theory. 't Hooft argued that the obstruction to coupling $G$ to classical background gauge fields is preserved under the Renormalization Group flow. The obstruction is usually referred to as an anomaly.

This idea has had powerful applications, especially when the associated obstruction and symmetry $G$ are continuous and the theory has  a weak coupling limit. This is due to the fact that the anomaly could be computed explicitly (for a review see~\HarveyIT).  Rather successful methods have been developed for computing continuous anomalies even for strongly coupled theories (especially for supersymmetric theories). Recently, there has been a lot of interest in discrete anomalies and their implications. There is indeed a rich landscape of theories that have discrete anomalies. Perhaps the simplest example is that of a free fermion in 2+1 dimensions, which has a parity anomaly~\refs{\RedlichDV,\RedlichKN}. For more sophisticated examples see, for example,~\refs{\SeibergRSG,\BeniniDUS} and references therein.

In addition, discrete anomalies have been recently used extensively in the context of topological insulators and Symmetry Protected Topological phases, in which the anomalies are canceled by coupling to a higher-dimensional theory (this is known as anomaly inflow~\CallanSA).
For a review of and references to applications in condensed matter systems see~\ChenPG.

It has been shown recently that discrete anomalies can arise even in the absence of fermionic degrees of freedom~\refs{\KapustinLWA,\KapustinZVA}.

In this note we study $4d$ $SU(N)$ Yang-Mills theory as a function of $\theta$.  (For early related discussion in two and four dimensions, see e.g.\ \refs{\DashenET\ColemanUZ\DAddaVBW\BaluniRF\WittenBC\WittenVV-\WittenSP}.) First, we review the fact that the system is not quite $2\pi$ periodic as a function of $\theta$.  The anomaly in shifting $\theta$ by $2\pi$ becomes clear either when we consider the system with a boundary or when we turn on nontrivial background fields for its center.  This discussion will lead us to a new time reversal anomaly at $\theta=\pi$. More precisely, we will show that there is a mixed anomaly involving time reversal and the global center 1-form symmetry. (We can use time reversal and CP interchangeably.) The center symmetry is a 1-form symmetry \refs{\KapustinGUA,\GaiottoKFA}, because the associated conserved charge is associated to a co-dimension 2 surface (which can wrap a loop). Standard 0-form symmetries are of course associated to co-dimension 1 surfaces (which wrap points).  We will review the results of~\refs{\KapustinGUA,\GaiottoKFA} that are needed to follow this paper.
While time-reversal is a symmetry of the theory, if we couple the center 1-form $\Z_N$  symmetry to classical background gauge fields, time reversal is broken by a $c$-number. In other words, we have a mixed time-reversal/center symmetry 't Hooft anomaly.

As always with 't Hooft anomalies, they do not invalidate the symmetry, but they lead to powerful constraints on the phases of the theory. In our case, assuming that the pure gauge $SU(N)$ theory is gapped for all $\theta$,\foot{This is widely believed to be the case, based on the results of lattice simulations, the analysis of supersymmetric Yang-Mills with softly broken supersymmetry, and holographic models (at large $N$). This assumption may well be false for small values of $N$, especially $N=2$. We will pay special attention to the case of $SU(2)$ gauge theory.} with a trivial non-degenerate vacuum at $\theta=0$, we derive  that time reversal is spontaneously broken at $\theta=\pi$. Therefore, the vacuum at $\theta=\pi$ is twofold degenerate. Furthermore, the anomaly has interesting implications for the phase diagram of the theory at finite temperature. One can ask at which temperature the CP symmetry is restored. The anomaly implies that it has to be restored at a temperature that is not lower than the deconfinement transition. Otherwise, some Wilson loops would not admit an area law. If the restoration of CP is at a temperature strictly higher than the deconfinement transition, then there is a new phase of the theory with $2N$ vacua (but the space-like Wilson loop is confined).

We will analyze in detail the case of $SU(2)$ gauge theory and make some proposals regarding the phase diagram.
In this particular case the deconfinement transition is second order and the possible phase diagrams are quite rich. In addition, there is no convincing argument that the theory is gapped at zero temperature around $\theta=\pi$, or that it is confined at finite temperature around $\theta=\pi$ (in the sense that space-like loops have an area law). We investigate these different scenarios in light of the constraints from anomaly matching.
We suggest two possibilities that are particular to $SU(2)$ gauge theory. One involves a new gapped phase with one ground state (with all the 0-form symmetries preserved) and  a perimeter law for space-like Wilson loops. The phase diagram would contain two critical points with $O(2)$ symmetry. This symmetry is ``accidental'' or ``hidden'', i.e.\ arises only at long distances. This is like a finite temperature Higgs phase. The second scenario is that the theory is in fact gapless already at {\it zero} temperature around $\theta=\pi$. In this case a natural proposal involves a four-dimensional Coulomb phase. (There could also be an interacting 4d CFT for special values of $\theta$.)

\subsec{An Analogous $2d$ System}

The anomaly that we described above and its consequences are reminiscent of the $2d$ $\C\P^{n-1}$ model.\foot{We thank E.~Witten for an extremely useful discussion about this analogy.} We can represent this model as $n$ complex scalars $z^i$ with the constraint $\sum|z^i|^2=1$, which are coupled to a $U(1)$ gauge field $a$ associated with their phase rotation.  The $U(1)$ field can have a $\theta$-term,
\eqn\thetat{{\theta\over 2\pi} \int da ~,}
which can be interpreted as as a background electric field.  The parameter $\theta$ seems to be $2\pi$ periodic.

Our discussion of this model will follow the lines of \BeniniDUS.  Naively, the global symmetry of the model is $SU(n)$.  But in fact, only $PSU(n) =SU(n)/\Z_n$ acts faithfully on the gauge invariant operators.  Therefore, it is useful to couple the system to background $PSU(n)$ gauge fields $A$.  More precisely, the gauge fields in the problem $\CA$ are $U(n)\cong (U(1)\times SU(n))/\Z_n$ fields.  Here the $U(1)$ field $a$ is dynamical and the $PSU(n)$ fields $A$ are classical.  The $\theta$-term \thetat\ is
\eqn\thetatg{{\theta\over 2\pi n} \int\Tr F_\CA ~,}
where $F_\CA$ is the $U(n)$ field strength.
Here we see that $\theta $ is $ 2\pi n$ periodic rather than $2\pi$ periodic.  More explicitly, for $A$ an $SU(n)$ background field (e.g.\ $A=0$) we recover \thetat\ with its $2\pi$ periodicity.  But when $A$ is a connection on a $PSU(n)$ bundle $E$ which cannot be lifted to an $SU(n)$ bundle, the expression \thetatg\ is more subtle.  In this case the dynamical field $a$ is not quite a $U(1)$ gauge field and $\int da$ is not a multiple of $2\pi$ -- it is correlated with the background field $A$.  For $\theta=2\pi k$ it is
\eqn\thetatgw{{2\pi k \over 2\pi n}\int \Tr F_\CA = {2\pi k\over n} \int w_2(E)~,}
where $w_2(E)$ is the second Stiefel-Whitney class of the $PSU(n)$ bundle $E$.  (On a closed  oriented 2-manifold $\int w_2(E)$ is an integer modulo $n$, and on an open manifold its integral is well-defined only if one specifies a trivialization of $E$ on the boundary.)  In other words, as we shift $\theta$ by $2\pi$ the local physics is unchanged, but a nontrivial term in the background fields \thetatgw\ is added to the Lagrangian.

This lack of $2\pi$ periodicity in the background fields has a simple physical interpretation.  As $\theta$ is increased from $0$ to $2\pi$ a pair of dynamical quanta  $z^i$ and $\bar z_i$ is created, separated, and moved to the boundary to screen the background electric field.  This is the physical reason for the apparent $2\pi$ periodicity.  But a closer inspection shows that the periodicity is not precise.  After the pair creation, the boundary of the system carries a nontrivial $SU(n)$ representation.  The whole system is still in a $PSU(n)$ representation.  But locally the boundary is not in a representation of the true global symmetry group $PSU(n)$.  The nontrivial term \thetatgw\ is a reflection of that phenomenon.  And its more subtle definition when the manifold has a boundary is associated with the $z^i$ particle there.

Stating it differently, if we start at $\theta=0$ without a $w_2$ term for the background field $A$ and we continuously change $\theta $ to $2\pi$, we end up with a nontrivial $w_2$ term corresponding to the nontrivial $SU(n)$ representation at the boundary.  This is similar to the phenomenon occurring in the famous Haldane chain.

This lack of $2\pi$ periodicity is particularly interesting at $\theta=\pi$.  At this point the system is naively time-reversal invariant.\foot{Equivalently, we can say that the system is $\CC\CT$ invariant for all $\theta$ and it violates $\CT$ and $\CC$ at generic $\theta$.  Then the symmetry we discuss here can be taken to be $\CC$ instead of $\CT$. We thank M.~Metlitski for a discussion.} The argument for that uses the fact that under time reversal $\theta \to -\theta$ and then using the $2\pi$ periodicity we find that $\theta =\pi$ is time reversal invariant.  But now we see that the system is not quite $2\pi$ periodic.  Instead, there is an anomaly in that transformation.  This means that for trivial $A$ the system at $\theta=\pi$ is time reversal invariant, but it is not invariant for nontrivial $A$.

This mixed anomaly between time reversal symmetry and the global $PSU(n)$ symmetry leads to 't Hooft anomaly conditions constraining the long distance behavior -- the IR system should reflect the same 't Hooft anomaly.  Indeed, for $n>2$ time reversal is spontaneously broken at $\theta =\pi$ with a two-fold degeneracy of the vacuum (see for instance~\AffleckTJ).  And for $n=2$ the system is gapless.

Our discussion of the $4d$ $SU(N)$ theory will be quite similar to that.  Instead of the global $PSU(n)$ symmetry, we will have a one-form global $\Z_N$ symmetry.  The periodicity of $\theta$ will be $2\pi N$ rather than $2\pi$ and the nontrivial 't Hooft anomaly will have consequences at $\theta=\pi$.

\subsec{Outline}

The outline of the paper is as follows. In section~2 we study some general aspects of $SU(N)$ Yang-Mills theory and derive the time reversal anomaly. We exhibit the consequences of this anomaly explicitly in the softly broken supersymmetric $SU(N)$ theory. In section~3 we begin our study of $SU(2)$ Yang-Mills theory on a circle, or equivalently at finite temperature. We discuss in detail the symmetries, the consequences of the new anomaly, the high temperature phase, the various domain walls,  and the confinement-deconfinement transition.
In section~4 we define a mixed gauge theory, which is essentially an orbifold of the original gauge theory on a circle. It has the advantage that the consequences of the anomaly are clearer in the orbifold theory, while no information about the original theory is lost. The mixed gauge theory has a $\bb D_8$ symmetry, and we discuss its various phases. We also describe phases of $SO(3)$ gauge theory at nonzero temperature.
In sections 5, 6 we discuss two possible scenarios for the phase diagram of the $SU(2)$ theory, assuming that the theory remains gapped at zero temperature for all theta. In section 5 we describe a scenario where
there is a phase with 8 vacua and a completely broken $\bb D_8$ symmetry. We argue that this scenario, appropriately modified, is in fact very natural for $SU(N)$ Yang-Mills theory at sufficiently large $N$. In this scenario, a general inequality is derived relating the temperatures at which CP and the center symmetry are restored.
 In section 6 we describe a situation with
a region of unbroken $\bb D_8$ symmetry.
We explain the consequences of these two phases for the original $SU(2)$ theory and make some detailed predictions about the phase diagrams. In section 7 we discuss  possible phase diagrams assuming a phase transition at zero temperature but non-zero $\theta$.  In the first three appendices we discuss an interesting modification of the topological $\Z_2$ gauge theory in two dimensions, provide another derivation of the $\bb D_8$ symmetry of the mixed gauge theory, and verify that this symmetry is free from 't Hooft anomaly. This is crucial for the consistency of the phase diagram with unbroken $\bb D_8$ symmetry. In the fourth appendix we present pedagogical quantum mechanical examples that exhibit anomalies.  These theories will appear on some of the domain walls that we encounter.

Let us comment on our notation. When discussing discrete 1-form symmetries and their gauging, it is convenient to adopt a lattice regularization of the gauge theory. We will use a slight modification of the Wilsonian lattice gauge theory where a triangulation is used in place of a hypercubic lattice. The advantage is that discrete gauge fields can be thought of as simplicial cochains, and an  action for them can be written in terms of standard operations such as the coboundary operator and the cup product. We will denote by $C^p(X,\Z_N)$ the Abelian group of $p$-cochains with values in $\Z_N$, and by $Z^p(X,\Z_N)$ its subgroup consisting of closed $p$-cochains (i.e.\ {\rm mod}\ $N$ $p$-cocycles). The coboundary operator on cochains is denoted $\delta$, to distinguish if from the exterior differential $d$ on differential forms. A $\Z_N$ gauge field is represented on the lattice by a $\Z_N$ 1-cocycle $a\in Z^1(X,\Z_N)$, and in the continuum by a $U(1)$ gauge field $A$ satisfying the constraint $NA=d\phi$, where $\phi$ is a $2\pi$-periodic scalar. The dictionary between $a$ and $A$ is, roughly,  $A={2\pi \over N} a$. Thus a Wilson loop in the charge-1 representation of $\Z_N$ along a closed loop $\Gamma$ is written as $\exp(i\int_\Gamma A)$ in the continuum and as $\exp({2\pi i\over N}\int_\Gamma a)$ on the lattice. When we write down an action for a $\Z_N$ gauge theory on a lattice, we impose the constraint $\delta a=0$ by means of a Lagrange multiplier field $b\in C^{n-2}(X,\Z_N)$, $n={\rm dim}\, X$, so that the action is
\eqn\actionnormalization{
{2\pi i\over N} \int b\cup \delta a.
}
The corresponding action in the continuum is
\eqn\actionnormalizationtwo{
{iN\over2\pi}\int B\wedge dA,
}
where $B$ is an $(n-2)$-form $U(1)$ gauge field and $A$ is an unconstrained 1-form $U(1)$ gauge field. The two actions are related by a formal substitution $B\to {2\pi\over N} b$, $A\to {2\pi\over N} a$, $\wedge\to\cup$, $d\to\delta$.

\newsec{$SU(N)$ Yang-Mills Theory in Four Dimensions}

\subsec{Symmetries}

We begin by considering $SU(N)$ Yang-Mills theory on $\IR^4$. The action is
\eqn\actionfour{S=\int d^4x \ {\rm Tr}\left( {-1\over 4 g^2}F\wedge \star F+{i\theta\over 8\pi^2}F\wedge F\right)~,}
with $F$ the field strength of the $SU(N)$ gauge field. The parameter $\theta$ is periodic
\eqn\periodic{\theta\sim\theta+2\pi~.}
While the theories with $\theta$ and $\theta+2\pi$ are equivalent (in the sense that there is a similarity transformation\foot{The similarity transformation is implemented by the unitary operator $U=e^{{i\over 4\pi} \int_{\Sigma_3} Tr \left(A\wedge dA+{2\over 3 }A^3\right)}$, where $\Sigma_3$ is the spatial slice.} that brings the Hamiltonian of the theory at $\theta$ to that of the theory at $\theta+2\pi$), the identification of line operators is non-trivial due to the Witten effect~\WittenEY. We will elaborate on this point below.
A CP transformation acts on $\theta$ as
\eqn\CP{{\rm CP} : \qquad \theta\to-\theta~.}
Therefore at $\theta=0$ and at $\theta=\pi$ the theory is CP-invariant. (Since CPT is always a symmetry, as we said, we  use CP and time reversal interchangeably.) Below we will argue  that at $\theta=0$ CP is preserved by the vacuum while at $\theta=\pi$ it is spontaneously broken (see e.g.~\refs{\DashenET,\CreutzXU} for a discussion of related questions). Therefore, at $\theta=\pi$ there are two gapped vacua and a domain wall. For other values of $\theta$, it is widely believed that there is a single gapped vacuum.
Hence, at $\theta=\pi$ there is a first order phase transition with a spontaneously broken CP symmetry.

In the rest of this subsection we review some useful facts about the $\Z_N$ 1-form global symmetry of our system \refs{\KapustinGUA,\GaiottoKFA}, which we will refer to as the the center symmetry.   Given a closed two-dimensional surface $\Sigma_2$ we can associate to it an operator $U(\Sigma_2)$ which is our $\Z_N$ charge generator. All local operators are neutral under this symmetry. Hence, this symmetry exists for all $\theta$.

In four dimensions, such a surface $\Sigma_2$ can have a nonzero linking number with a loop $\gamma$. If we take the Wilson line in the fundamental representation $W_F=Tr_FPe^{i\int_\gamma A}$ and wrap a homologically trivial $\Sigma_2$ around $\gamma$ so that the linking number is $1$,  then we find
\eqn\WilsonLine{
\langle U(\Sigma_2) W_F\cdots\rangle =w\langle W_F\cdots\rangle~,}
with $w=e^{2\pi i / N}$ an $N$th root of unity. Equivalently, in the Hamiltonian language, the operators $U(\Sigma_2)$ and $W_F(\gamma)$ satisfy
\eqn\USigma{
U(\Sigma_2) W_F(\gamma) U(\Sigma_2)^{-1} =w^{(\Sigma_2,\gamma)}W_F(\gamma)~,}
where it is assumed that both $\Sigma_2$ and $\gamma$ lie in the spatial slice $\Sigma_3$, and  $(\Sigma_2,\gamma)$ is their intersection number in $\Sigma_3$.

Let us discuss the other line operators in the theory. We use the notation of~\AharonyHDA\ and organize line operators into families labeled by pairs $(a,b)\in \Z_N\times\Z_N$ . The fundamental Wilson line belongs to the  $(1,0)$ family. More generally, a Wilson line in a tensor representation of $SU(N)$ with $k$ fundamental and $l$ anti-fundamental indices belongs to the $(k-l,0)$ family. Line operators in the family $(a,b)$ have magnetic charge $b$ and for $b\neq 0$ must be attached to a topological surface. This topological surface is made of $b$ copies of $U(\Sigma_2)$ such that $\partial\Sigma_2$ is the support of the line  operator $(a,b)$. The charge of lines with  $b\neq 0\ {\rm mod}\ N$  under the center symmetry is not well defined since there could be counter-terms when the two surfaces intersect. But for $b=0\ {\rm mod}\ N$ the charge is well defined according to~\WilsonLine.

The 1-form $\Z_N$ symmetry does not suffer from an 't Hooft anomaly. That is, the theory can be coupled to a background $\Z_N$ 2-form gauge field $B$ in such a way that the partition function is invariant under $\Z_N$ 1-form gauge transformation. This is so because the usual Wilsonian lattice regularization of $SU(N)$ Yang-Mills theory has a manifest 1-form $\Z_N$ symmetry which acts locally \KapustinGUA. On the lattice, $B$ is represented by a 2-cocycle with values in $\Z_N$, and 1-form gauge symmetry acts on it by $B\mapsto B+\delta\lambda$, where $\lambda$ is a 1-cochain with values in $\Z_N$. Coupling the theory to $B$ can also be thought of as inserting a surface operator $U(\Sigma_2)$ such that the homology class of $\Sigma_2$ modulo $N$ is Poincar\'{e} dual to $[B]\in H^2(X,\Z_N)$. The choice of $\Sigma_2$ within its homology class is immaterial, precisely because the 't Hooft anomaly is absent.

If all $(a,0)$ line operators with $a\neq 0$ are confined (i.e.\ their expectation values have an area law), the 1-form $\Z_N$ center symmetry is unbroken \GaiottoKFA. It is believed that $SU(N)$ gauge theory without matter confines for all values of $\theta$ and hence the 1-form $\Z_N$ symmetry is always unbroken at zero temperature. In more general $SU(N)$ gauge theories it might happen that for some nonzero $a$, which is a divisor or $N$, the lines $(a k ,0)$ with integer $k$ have a perimeter law, while other lines have area law \CachazoZK.  In that case only a $\Z_a \subset \Z_N$  subgroup is unbroken \GaiottoKFA.  This is the subgroup that leaves the line $(a,0)$ invariant.

\subsec{A CP Anomaly for Even $N$}

As we have remarked above, the identification~\periodic\ involves a nontrivial action on the line operators in the theory~\WittenEY. Indeed, the map $\theta\to\theta+2\pi$ induces a transformation
\eqn\mapWi{\theta\to\theta+2\pi: \qquad (a,b)\to (a+b,b)~.}
An immediate consistency check is that lines that are attached to a topological surface remain such under this transformation and vice versa. In particular, if $b=0\ {\rm mod}\ N$ then $a \ {\rm mod}\ N$ is invariant under the transformation and hence also the charge under the center symmetry is invariant. But ~\mapWi\ has consequences for other line defects, which do not have a well-defined $\Z_N$ charge.

Consider for instance the loop $(0,1)$ which is attached to our $\Z_N$ surface. It transforms to the loop $(1,1)$, which is also attached to a surface. Now consider coupling  the theory to a $\Z_N$ two-form {\it background} gauge field $B$. This is equivalent to creating a network of surface operators $U(\Sigma_2)$ such that the homology class of $\Sigma_2$ is Poincar\'{e} dual to $[B]\in H^2(X,\Z_N)$.  In particular, we can consider a homologically trivial $\Sigma_2$, which links the $(0,1)$ line defect once, and thus has intersection number $1$ with the topological surface attached to the line defect.

But there is a certain ambiguity in the gauging procedure, due to the possibility of adding counter-terms depending only on $B$ (one may think of this counter-term as a seagull term that we are free to add). Geometrically, this ambiguity arises from the freedom to assign weights to intersection points of surface operators . We can choose the counter-terms for $B$ to vanish, namely, when two $\Sigma_2$ surfaces intersect (at a point), we do not add any weight. Then the intersection of $\Sigma_2$ with the surface attached to the $(0,1)$ line operator does not introduce any weight and indeed $\Sigma_2$ can be shrunk away.  We can see however that the transformation under $\theta\to\theta+2\pi$, taking the loop $(0,1)$ to the loop $(1,1)$, loosely speaking, changes the charge under the 1-form symmetry. Indeed, $(1,1)$ operator can be thought of as a composite of a $(0,1)$ operator and a $(1,0)$ operator, and the latter gives an extra factor $w$ when one tries to shrink away $\Sigma_2$. To get rid of this factor, one needs to postulate that the transformation $\theta\to \theta+2\pi$ induces a counter-term in the action. For even $N$, this counter-term takes the form
\eqn\anomaly{\theta\to\theta+2\pi : \qquad \Delta S ={2\pi i(N-1)\over 2N} \int_X B \cup B ~.}
(The derivation of this formula will be presented in subsection~2.4.) The case of odd $N$ is considered in the next subsection. This counter-term guarantees that when two $\Sigma_2$ surfaces intersect, we pick up a phase $e^{-2\pi i /N}$. This is necessary for the consistency of the mapping~\mapWi. More generally, if the counter-term was there in the beginning, it is shifted by the amount \anomaly\ when we shift $\theta\to\theta+2\pi$.

It is a somewhat subtle fact that the expression \anomaly\ is well-defined (for even $N$). Indeed, if we regard $B$ as integer-valued 2-cocycle, then replacing $B\to B+N b$ for some integral 2-cochain $b$ changes $\Delta S$ by
\eqn\anomalychange{
-{2\pi i N\over 2} \int_X b\cup b-{2\pi i \over 2}\int_X (b\cup B+B\cup b).
}
Here we used the fact that $N$ is even. The first term is an integral multiple of $2\pi i$ for even $N$ and thus is trivial when exponentiated. The second term is either an integral or a half-integral multiple of $2\pi i$, because the cup product is not commutative on the cochain level. It turns out one can correct the expression \anomaly\ to make sure that $\Delta S$ is always shifted by an integral multiple of $2\pi i$ ~\KapustinGUA . The corrected expression uses the Pontryagin square instead of the usual cup square.

The fact that $\theta\to \theta+2\pi$ induces a counter-term for the 2-form background gauge field has many important consequences, which we will now explain.

It is necessary to add the counter-term~\anomaly\ in order to guarantee that $\theta$ and $\theta+2\pi$ are equivalent. However, if we like to make $B$ a dynamical gauge field then we have to pick a particular counter-term. But then $\theta\to \theta+2\pi$ cannot describe equivalent theories. Indeed, if we turn $B$ into a dynamical gauge field, we find the $PSU(N)$ gauge theory, where $\theta$ and $\theta+2\pi$ are not equivalent. The periodicity of $\theta$ in $PSU(N)$ gauge theory is, instead, $4\pi N$ (on general non-spin manifolds and for even $N$). In particular, the $PSU(N)$ theory at $\theta=\pi$ does not have CP symmetry.

Most interesting is to consider the implications of the counter-term~\anomaly\ at  $\theta=\pi$, where $CP$ is a symmetry in the $SU(N)$ theory. We see that a CP transformation has to be accompanied by adding the counter-term~\anomaly.\foot{For $\theta=0$ CP symmetry does not need to be accompanied by adding a counter-term. There is therefore no anomaly.}
This means that there is a mixed 't Hooft anomaly involving CP and the 1-form $ \Z_N$ symmetry.
The discussion in the previous paragraph shows one manifestation of this mixed anomaly: if we gauge the 1-form $\Z_N$ symmetry, then the CP symmetry is explicitly broken.\foot{If the 2-form $\Z_N$ gauge field is not dynamical but it has some nontrivial value, then the CP symmetry is still present but the vacuum transforms nontrivially under a CP transformation. This can be interpreted by saying that with a nontrivial background $B$ we have introduced the 't Hooft twisted boundary conditions defect~\tHooftUJ, which carries charge under CP. Indeed, nontrivial values of the 2-form $\Z_N$ gauge field are classified by same cohomology group as the 't Hooft twisted boundary conditions. In $PSU(N)$ we have to sum over all such defects and hence the CP symmetry is destroyed. } This is a standard phenomenon in situations with a mixed 't Hooft anomaly -- if we gauge one symmetry the other is explicitly broken.

There is another striking implication that we can derive by anomaly matching: if we assume that at $\theta=\pi$ the theory is confined (and hence the 1-form symmetry is preserved), the CP symmetry is either spontaneously broken or the vacuum supports a nontrivial theory at long distances\foot{This nontrivial theory could  be either a gapless theory or a topological field theory.} matching the anomaly. But it is well-known that at $\theta=0$ the theory has a trivial gapped vacuum. Hence either the CP symmetry is spontaneously broken at $\theta=\pi$, or the gap vanishes for some $\theta$ in the interval $(0,\pi]$, or there is a 1st order phase transition to a topologically-ordered state for some $\theta$ in the interval $(0,\pi)$.

The anomaly~\anomaly\ is reminiscent of the more familiar parity anomaly in three dimensions~\refs{\RedlichDV,\RedlichKN}. There, a time-reversal transformation induces a properly-quantized Chern-Simons counter-term for the background gauge field. The anomaly can be interpreted as a mixed time-reversal/$U(1)$ anomaly. Here we see that $SU(N)$ Yang-Mills theory has a similar mixed time-reversal/$\Z_N$ center anomaly.

Let us make this analogy a little more precise. Here we discuss $SU(N)$ for even $N$, and in the next subsection we extend the discussion to odd $N$.

We should require from the theory at $\theta=\pi$ to be invariant under CP and all of its counter-terms to be invariant as well. Let us start with the the most general counter-term, namely\foot{This is a bit schematic. As explained above, one should really use Pontryagin square instead of the cup square.}
\eqn\properlyeven{
{2\pi i p \over 2N} \int_X B \cup B
}
where $p$ is defined modulo $2N$ \KapustinGUA .
A CP transformation at $\theta=\pi$  flips the sign of $p$ and adds the term~\anomaly . In other words,  it acts on $p$ as follows:
\eqn\ptransformationeven{
p\to -p+N-1 ~.
}
Since gauge invariance forces $p$ to be quantized, we cannot choose the counter-terms to preserve CP.  Indeed, there is no integral fixed point to the transformation~\ptransformationeven.
This is exactly reminiscent of the famous time reversal anomaly of a free Dirac fermion in three dimensions. There, a time reversal transformation shifts the action by the classical term ${i\over 4\pi}\int A\wedge dA$. Since there is no properly quantized $3d$ counter-term that can cancel this effect, there is a mixed anomaly that involves time-reversal and the $U(1)$ symmetry of the free fermion.  The only difference between this $3d$ example and our $4d$ problem is that the ordinary $U(1)$ symmetry of the $3d$ problem is replaced by a $\Z_N$ one-form global symmetry in the $4d$ problem.

Note that for $\theta=0$ the CP transformation acts by $p\to -p$, so one can achieve CP invariance by not including any counter-terms. This reflects the fact that there is no mixed anomaly for $\theta=0$.

\subsec{A CP Anomaly for Odd $N$}

For odd $N$ a properly quantized counter-term has the form
 \eqn\properlyodd{{2\pi i p \over 2N} \int_X B \cup B ~,}
 with  an even $p$ \KapustinGUA. Only $p$ mod $2N$ is meaningful.  If we take $\theta=\pi$ then a CP transformation shifts the counter-term by
\eqn\anomalyodd{\Delta S ={2\pi i (N-1)\over 2N} \int_X B \cup B ~.}
This ensures that the intersection of two surface operators picks an extra factor $w^{-1}$ after the shift.
Therefore the coefficient of the counter-term transforms as follows under CP:
\eqn\ptransformationodd{
p\longrightarrow -p+ N-1~.
}
This transformation has exactly one fixed point with even $p$ in the range $[0,2N-1]$: $p=(N-1)/2$ if $N=1\ {\rm mod}\ 4$ and $p=N+(N-1)/2$ if $N=3\ {\rm mod}\ 4$. Thus for any odd $N$ there exists a choice of the counter-term that preserves both the 1-form $\Z_N$ symmetry and CP at $\theta=\pi$. Therefore there is no mixed anomaly for odd $N$.

Note however that at $\theta=0$ CP transformation acts by $p\to -p$. For odd $N$, this transformation has a unique fixed point, namely $p=0$. Therefore, for odd $N$, even though there exist separate choices that preserve CP at $\theta=0$ and $\theta=\pi$,  there is no consistent choice  that preserves CP invariance for both $\theta=0$ and $\theta=\pi$ simultaneously. This is almost as good as saying that there is an anomaly. Indeed, it implies that there cannot be a  regularization which is CP invariant for both $\theta=0$ and $\theta=\pi$. Similarly, the infrared consequences of this slightly weaker statement for odd $N$ are similar to the 't Hooft anomaly matching conditions. Indeed, the infrared theories at $\theta=0$ and $\theta=\pi$ cannot be both trivial with unbroken CP invariance.

Lattice simulations indicate that CP is not spontaneously broken at $\theta=0$. Therefore, assuming that the theory is gapped and confining at $\theta=\pi$, it must either break CP spontaneously, or be in a topologically ordered state with both a CP symmetry and a 1-form $\Z_N$ symmetry. The former option is much more probable, certainly for sufficiently large $N$. In the case of a topologically ordered state with both a CP symmetry and a 1-form $\Z_N$ symmetry, a phase transition must occur for $\theta$ in the interval $(0,\pi)$.

\subsec{A Continuum Description of the CP Anomaly}

It is useful to present the above anomaly~\anomaly\ from the continuum point of view. In~\anomaly\ we have used a cup product between $\Z_N$ valued 2-cocycles, but here we will show that it can be also understood in the continuum by embedding the 1-form discrete symmetry above into a continuous symmetry. This discussion uses some results that appeared in~\refs{\KapustinGUA,\GaiottoKFA}.

In the continuum a $\Z_N$ 2-form gauge field is represented by a pair $(B,C)$, where $B$ is a $U(1)$ 2-form gauge field, $C$ is a $U(1)$ 1-form gauge field, and they satisfy a relation\foot{It is tempting to try to solve for $B$ and work with $C$ alone. But one must resist this temptation,  since $B$ is not a 2-form, but a 2-form gauge field, and $NB$ does not completely determine $B$.} $NB=dC$. This relation ensures that $e^{i\oint _\Sigma B}$ is an $N^{\rm th}$ root of unity for any 2-cycle $\Sigma$.  Moreover, given a 2-chain $\Sigma$ whose boundary is $N$ times a 1-chain $\Gamma$, one can define a more general observable  with the same property:
\eqn\Nsurface{
U(\Sigma,\Gamma)=\exp\left(i\int_\Sigma B\right)\exp\left(-i\int_\Gamma C\right).
}
The relation between the discrete field $B_{discr}\in Z^2(X,\Z_N)$ and the continuum fields $B,C$ is, roughly, $B={2\pi \over N} B_{discr}$.
A $\Z_N$ 1-form gauge symmetry transformation is replaced by a pair $(\lambda,f)$, where $\lambda$ is a $U(1)$ gauge field and $f$ is a periodic scalar (i.e.\ $f$ is a continuous function on $X$ with values in $\R/2\pi\Z$). These transformations act as follows:
\eqn\contoneform{
B\to B+d\lambda,\quad C\to C+df+N\lambda.
}
This symmetry leaves invariant the observables \Nsurface\  as well as the constraint $NB=dC$.

We extend the original $SU(N)$ gauge field $a$ to a $U(N)$  gauge field $a'$. We add a dynamical 2-form $U(1)$ gauge field $u$ and a background  pair $(B,C)$ as above such that our Lagrangian now has the term
\eqn\addtoLM{{1\over 2\pi} u \wedge (\Tr F' - dC)~,}
where $ F'$  is the field strengths of $a'$.  $u$ acts as a Lagrange multiplier setting $\Tr a' \sim C$ up to a gauge transformation.  Clearly, for trivial $(B,C)$ we find the original $SU(N)$ gauge theory. Under the gauge symmetry \contoneform\ the $U(N)$ gauge field $a$ transforms as follows:
\eqn\contatransf{
a\to a+ \lambda\unit.
}
where $\unit$ a unit matrix in $U(N)$. Since $\lambda$ is arbitrary, we can use this gauge freedom to reduce a $U(N)$ gauge field to an $SU(N)$ gauge field.

Next, we would like to replace an action for the $SU(N)$ field $a$ by an action for the $U(N)$ field $a'$.  In order to preserve the one-form gauge symmetry \contoneform, we replace $F\to F'-B\unit$.  For example, the instanton number density becomes
\eqn\addtheta{\eqalign{
{1\over 8\pi^2}  \Tr (F\wedge F) \qquad\longrightarrow\qquad &  {1\over 8\pi^2} \Tr (F'-B\unit)\wedge (F' -B\unit) \cr
&={1\over 8\pi^2} \Tr (F'\wedge F') -{1\over 4\pi^2N} dC\wedge \Tr F'+{1\over 8\pi^2N} dC\wedge dC}}
where $F $ is the field strengths of $a$. Taking into account the constraint $\Tr F'=dC$, the theta-term becomes
\eqn\newtheta{
{i\theta\over 8\pi^2}\int_X \left(\Tr (F'\wedge F')-{1\over N} dC\wedge dC\right).
}

Finally, we can also add to the Yang-Mills action a gauge-invariant counter-term for the background field $(B,C)$. The most general counter-term is
\eqn\addct{{i pN\over 4\pi} \int_X B\wedge B = {i p\over 4\pi N}\int_X dC \wedge dC~.}
Here $p$ must be integer for even $N$ and an even integer for odd $N$. In both cases we have an identification $p\sim p+2N$ \KapustinGUA . One can easily check that \addct\ is invariant under \contoneform.

Let us examine the behavior of the action under $\theta \to \theta +2\pi$.  Note first that for a $U(N)$ gauge field $a'$,  $\int_X \Tr (F'\wedge F')/8\pi^2$ can be half-integral, but the expression
\eqn\ctwo{
 {1\over 8\pi^2} \int_X \left( \Tr (F'\wedge F') - \Tr F'\wedge \Tr F'\right)= {1\over 8\pi^2} \int_X \left( \Tr (F'\wedge F' )- dC\wedge dC \right)
}
is always integral. This is because the expression \ctwo\ is minus the second Chern number $c_2$ of the $U(N)$ bundle.
Thus the change in the theta-term \newtheta\ under $\theta\to\theta+2\pi$ is
\eqn\shifttheta{
-2\pi i c_2+{i(N-1)\over 4\pi N} \int_X dC\wedge dC.
}
Since $c_2$ is integral, the first term does not affect $\exp(S)$ and can be dropped. The second term can be absorbed into to a shift of the coefficient of the counter-term
\eqn\pshift{
p\to p+N-1.
}
Note that for $N$ odd it preserves the requirement that $p$ is even.

When we consider CP transformation at $\theta=\pi$, we get that $p$ transforms as
\eqn\CPpshift{
p\to -p+N-1.
}
This transformation has no integral fixed points for even $N$ and has a unique even fixed point for odd $N$. Thus we recover the analysis of the previous section.

When the periods of $dC$ are in $2\pi N \Z$, i.e.\ $B$ is trivial, we can be set $C$ to zero using the one-form gauge symmetry and then $a'$ is traceless (up to a gauge transformation).  This leads to the original $SU(N)$ theory.  Otherwise, the periods of $B $ are nontrivial and we find $PSU(N)$ bundles that are not $SU(N)$ bundles. If we make the two-form gauge field $B$ dynamical, then
$p$ is interpreted as a coupling constant in the Lagrangian, which determines how we sum over the various bundles. If $B$ is a background field then $p$ is interpreted as a counter-term.

Note that in our analysis thus far $p$ has been a counter-term in the ultraviolet. Then, it has to be properly quantized. But there is another possible application of the discussion above. If we have a theory with $\Z_N$ 1-form symmetry with a nontrivial ground state (gapless, or topologically ordered), it may lead to an {\it effective} fractional $p_{ir}$. We can then think of $p_{ir}$ as an intrinsic observable of the infrared theory, defined modulo the counter-term~\addct.
It would be interesting to study this observable. If the ground state is trivial, then $p_{ir}$ would have to be properly quantized. But the difference between the ultraviolet and infrared $p$'s is still a meaningful observable associated to the RG flow. This point of view is analogous to the Hall conductivity in three dimensions, which is an intrinsic infrared observable that is defined modulo an integer, see~\ClossetVP.

\subsec{Softly Broken ${\cal{N}}=1$ SYM}

The anomalies described above follow just from the existence of the center 1-form symmetry and CP symmetry. Hence, they should be present also in softly broken ${\cal{N}}=1$ SYM.
The supersymmetric theory with gauge group $SU(N)$ has $N$ vacua~\WittenDF. We define\foot{We use the notations and conventions of \IntriligatorAU.} the instanton parameter $\eta=\Lambda^{3N}=\mu^{3N} e^{-{8\pi^2\over g(\mu)^2} +i\theta}$, where $\mu$ is a real renormalization point and $\Lambda$ is complex. Then the gaugino condensate is
\eqn\guagc{\langle \lambda\lambda\rangle= \eta^{1/N}~.}
The $N$ roots of $\eta$ label $N$ different vacua.

Next, we softly break supersymmetry by adding to the Lagrangian a mass term for $\lambda$, $\delta {\cal L}=m\lambda^2+c.c.$, and first take $|m|\ll |\Lambda|$. The above vacua are no longer degenerate. The induced potential is just
\eqn\potinduced{\delta V= -m\langle\lambda^2\rangle+c.c.= -{\rm Re}(2m\eta^{1/N})~.}

Using the chiral anomaly the theory depends only on
\eqn\shiftt{\theta_{eff}=\theta + N {\rm Arg}(m)~}
and not separately on $\theta$ and ${\rm Arg}(m)$.
Therefore, without loss of generality we let $m$ be real and positive.  Let us assume that the physics changes smoothly as a function of $m$. Therefore, the
qualitative behavior of the low $m$ theory is the same as the behavior of the theory at  large $m$. This allows us to make contact with
the non-supersymmetric $SU(N)$ gauge theory with the same $\theta$-parameter.

For generic $\theta$ only one vacuum remains due to the potential~\potinduced. However for $\theta=\pi$ the situation is more interesting.
At $\theta=\pi$~\potinduced\ leaves two degenerate vacua for both even and odd $N$. This is precisely consistent with the anomaly prediction. Furthermore, the anomaly implies that this degeneracy is not lifted for any $m$. A similar analysis of the softly broken ${\cal N}=2$ theory can be found in~\KonishiIZ. Some aspects of softly broken supersymmetric QCD at large $N$ are discussed in~\DineSGQ.

Finally, as discussed before, it may be that as we increase $|m|$ the two vacua approach each other and lead to a nontrivial gapless ground state.  Indeed, the supersymmetric $2d$ $O(3)$ model is gapped and has two ground states for every $\theta$.  But as we break supersymmetry and flow to the non-supersymmetric model, the system is gapless at $\theta=\pi$.

Other interesting continuous deformations of Yang-Mills theory were considered in detail both at zero and nonzero temperature, see e.g.~\refs{\PoppitzNZ,\UnsalZJ}. In these models spontaneous breaking of CP symmetry is observed at $\theta=\pi$ and zero temperature.

\subsec{Boundaries, Interfaces, and Domain Walls}

Since the lack of $2\pi$ periodicity is crucial to our discussion we would like to point out another consequence of it, which is associated with an interface between a state with $\theta$ and a state with $\theta+2\pi$. Such an interface can arise dynamically as a domain wall between two vacua,
such as the two expected CP-breaking vacua at $\theta = \pi$ or between two supersymmetric vacua of an
$\CN=1$ supersymmetric theory.  The discussion of this interface in \refs{\DieriglXTA,\GaiottoKFA} stressed an anomaly in-flow mechanism and proposed a $U(1)_N$ topological field theory on the domain wall.  (A similar TQFT had been discussed earlier in the context of dynamical domain walls in $\CN=1$ supersymmetric theories in \AcharyaDZ.)

The anomaly inflow alone, of course, does not completely determine the dynamical theory on the interface. The $U(1)_N$ proposal is just one possible solution to the anomaly inflow constraint. Note in particular that $U(1)_N$ is a spin-TFT when $N$ is odd.
This was not a problem for the $\CN=1$ supersymmetric theory, but is unsuitable for the domain wall expected between the
two CP-breaking vacua of pure $SU(N)$ gauge theory at $\theta = \pi$.

Level-rank duality provides a naive identification between $U(1)_N$ and $SU(N)_{-1}$ Chern-Simons theories.
This identification, though, is precise only for fermionic theories: the quasi-particles of the two theories match
only when dressed by transparent fermion lines \HsinBLU. In particular, $SU(N)_{-1}$ is always a standard TFT not requiring a choice of spin structure.
It is natural to conjecture that the domain wall in the non-supersymmetric $SU(N)$ gauge theory
supports at low energy an $SU(N)_{-1}$ TFT.

An alternative way to produce such interfaces is to consider a non-dynamical modification of the UV theory. For example, we could
simply consider a position-dependent $\theta(x)$ with a sharp step at the interface, increasing from
$\theta_0$ to $\theta_0 + 2 \pi k$. This definition of the interface is completely equivalent to  adding to the bulk action (with constant $\theta$) an explicit
$SU(N)_k$ Chern-Simons action supported at the interface.

We may wonder what low energy degrees of freedom will appear at the interface. As the bulk is gapped,
the degrees of freedom can be expressed as a 3d theory, $T_k$, equipped with an anomalous $\Z_N$ 1-form symmetry,
which matches the anomaly of the UV Chern-Simons action.

The obvious candidate for $T_k$ is an $SU(N)_k$ Chern-Simons TFT. This is not the only possible candidate.
For example, if we were to smear the $2\pi k$ variation of $\theta$ in the UV over a distance
larger than the strong coupling scale, we would likely obtain a low energy theory of the form
$\left(SU(N)_1\right)^k$.

Indeed, if $T_k = SU(N)_k$, there must be a non-trivial phase transition as a function of the UV interface thickness.
This phase transition may be first or second order.  We can make a crude model for this transition. Consider a 3d quiver
$[ SU(N)_1 ]^k$ theory with bi-fundamental scalars between consecutive nodes. If the scalars are massive, we get an $[ SU(N)_1 ]^k$ theory in the IR. However, in the Higgs phase we get a $SU(N)_k$
TFT in the IR. Whether the transition is first or second order may depend on the couplings in the Lagrangian.

\newsec{$SU(2)$ Yang-Mills Theory on $Y \times S^1$}

We now study the $SU(2)$ theory at temperature $T={1\over \beta}$ by putting the theory on $Y \times S^1$, where $Y$ is a 3-manifold (compact or non-compact, as the case may be), and  $S^1$ has circumference $\beta$. Note that most of the discussion in this section (omitting subsection 3.4) is valid, after some appropriate adjustments, for $SU(N)$ gauge theory.  An important element in the analysis is the Polyakov loop, which measures the holonomy around the $S^1$
\eqn\Polyakov{U = P e^{i\oint_{S^1} A} \equiv e^{ i\beta \Phi}~,}
with $U$ a $2\times 2$ unitary matrix and $\Phi$ a $2\times 2$ hermitian matrix both depending on the coordinates on $Y$.

The first natural question to ask when studying the theory on $Y\times S^1$ concerns  the phases of the theory at long distances, much longer than $\beta$. If $Y$ is compact, this means that the diameter of $Y$ is much larger than $\beta$. This is described by an effective three-dimensional theory on $Y$.
The symmetries of the theory are inferred by dimensional reduction from four dimensions. On $X$  we had a 1-form $\Z_2$ symmetry for all $\theta$ and we had a CP symmetry at $\theta=0,\pi$. Upon reduction to three dimensions we find
\eqn\threedsymm{\matrix{ &  0-{\rm form} & 1-{\rm form} & {\rm space-time}\cr  \theta\neq 0,\pi   & {\Z_2} & {\Z_2} & T\cr \theta=0,\pi  & {\Z_2}\times {\Z_2} & {\Z_2} & T\cr } ~.}
The four-dimensional 1-form symmetry splits upon reduction to three dimensions to a 1-form and 0-form symmetry. At $\theta=0,\pi$ the four-dimensional CP symmetry becomes a global symmetry from the point of view of the three-dimensional theory, since it can be taken to act on the $S^1$. Finally, since CPT is preserved for all $\theta$, even when $CP$ is broken, time reversal must be a symmetry of the three-dimensional theory.

In 4d we had classes of line operators labeled by $(a,b)\in \Z_2\times\Z_2$. These line operators were genuine line operators for $b=0$, while for $b=1$  they required a topological surface operator attached to them. From the viewpoint of the three-dimensional theory, a genuine loop wrapping the $S^1$ becomes a local operator. A wrapped loop that is attached to a topological surface becomes an operator attached to a topological line. Loops that do not wrap the $S^1$ remain loops from the viewpoint of the three-dimensional theory. We often refer to such loops as ``space-like'' loops. They are labeled, as before, by $(a,b)$ with $a,b$ defined mod 2.
For example, the wrapped $(0,1)$ is attached to a topological line that generates the $\bb Z_2$ 1-form symmetry in three dimensions. The space-like (0,1) loop is attached to a topological surface that generates the 0-form $\bb Z_2$ symmetry in three dimensions.

\subsec{The CP Anomaly Reduced to Three Dimensions}

We have seen above that in the four-dimensional theory a $CP$ transformation at $\theta=\pi$ adds to the action the counter-term (in our analysis below we restrict to spin manifolds)
\eqn\sectwoa{\Delta S={i\pi \over 2}\int_X B\cup B, \quad B\in Z^2(X,\Z_2) ~.}
From the point of view of the three-dimensional theory at distances much larger than $\beta$, we have to split the background 2-form gauge field $B$ in 4d into a  2-form gauge field and a 1-form gauge field in 3d. They are gauge fields for the 1-form $\Z_2$ center symmetry and the 0-form $\Z_2$ center symmetry, respectively. We will call them $B$ and $A$, so that $B\in Z^2(Y,\Z_2)$ and $A\in Z^1(Y,\Z_2)$. Since from now on we work exclusively on $Y\times S^1$, this should not cause confusion.

The anomalies of the four-dimensional theory should be matched by the three-dimensional theory.\foot{This is true in the sense that the variation of the functional integral should be the same as that of the four-dimensional theory. Such anomaly matching across dimensions has appeared recently in the context of hydrodynamics (see e.g.~\refs{\LandsteinerCP\BanerjeeIZ-\JensenKJ}) and supersymmetry (see e.g.~\refs{\BonettiELA\DiPietroBCA-\ArdehaliHYA}).} Therefore, to reproduce~\sectwoa\ also the three-dimensional theory must acquire some counter-terms upon a CP transformation at $\theta=\pi$. This is obtained by dimensional reduction of~\sectwoa. Therefore, under a CP transformation at $\theta=\pi$ we must have
\eqn\sectwoa{\Delta S_{3d}=i\pi \int_Y A\cup B~.}

Since from the point of view of the three-dimensional theory CP is just a global $\bb Z_2$ symmetry,
we can easily couple a classical $\Z_2$ gauge field to it.\foot{From the four-dimensional point of view, that means considering
possibly non-orientable circle fibrations over the three-dimensional space-time.}
Let $a\in Z^1(Y,\Z_2)$ be the corresponding 3d 1-form background gauge field.
Then, the anomaly~\sectwoa\ can be viewed as arising by anomaly inflow from \eqn\inflow{S_{{\rm inflow}}=i\pi \int_M a\cup A\cup B~,}
where $a$ is extended to the auxiliary four-dimensional space $M$ with boundary $Y$ (which is not to be confused with the four-dimensional space $X=Y\times S^1$ where the Yang-Mills theory was defined) as a standard 1-form gauge field.
This can be viewed as a variation of the Dijkgraaf-Witten~\DijkgraafPZ\ construction, which includes a 2-form gauge field.

The anomaly~\inflow\ forbids the three symmetries, i.e.\ CP, center 0-form and center 1-form, to be simultaneously unbroken in a trivial gapped vacuum. This is therefore a mixed anomaly in three dimensions which involves two $\Z_2$ 0-form symmetries and one $\Z_2$ 1-form symmetry.

As we will see, this anomaly has profound implications for the phase diagram of the theory at finite temperature.

\subsec{The High Temperature Phases}

When the $S^1$ is very small compared to the dynamically generated scale in Yang-Mills theory, we can study the three-dimensional theory simply by dimensional reduction. This is the standard high-temperature expansion in Yang-Mills theory~\refs{\AppelquistVG,\GrossBR}. The three-dimensional theory has an $SU(2)$ gauge field with gauge coupling $g_{3d}^2\sim g_{YM}^2 \beta^{-1}$ and an adjoint scalar $\Phi$, arising as in \Polyakov.  It is convenient to pick a gauge where $\Phi=\phi(x)\left(\matrix{1 & 0 \cr 0& -1}\right)$. It is clear from \Polyakov\ that $\phi$ is subject to the identifications
\eqn\ident{\phi \sim \phi+{2\pi \over \beta} \sim -\phi~.   }
In the  $\beta\to 0$ limit the scalar $\phi$ becomes non-compact.
Equivalently, we can say that only the trace of the holonomy
\eqn\trholon{\Tr\ U = \Tr\ e^{ i\beta \Phi} = e^{i \beta\phi}+e^{-i \beta \phi}}
is well-defined.

In addition to the standard kinetic terms we have the interaction that is induced by the four-dimensional $\theta$ term:
\eqn\threedtheta{\delta S \sim \theta\int d^3 x \epsilon^{\mu\nu\rho}\Tr\left( D_\rho \Phi\ F_{\mu\nu}\right)~.}
If the theory is purely three-dimensional and $\Phi$ is a non-compact Lie-algebra valued section then such a term in the Lagrangian is trivial. This is because we can integrate by parts. But in the context of this theory arising from a circle compactification we have the identification~\ident\ and hence we are not allowed to integrate by parts since $\Phi$ is not a well-defined adjoint-valued scalar field.

We can now identify the symmetries of the problem. The center symmetry acts as
\eqn\symmetries{{\rm center}:\ \ \phi\to \phi+{\pi \over \beta}~ ,}
which is equivalent to its action on the holonomy $U\to -U$.
The 1-form symmetries only act on lines, not on local operators.
Time reversal acts by combining a reflection in one of the coordinates in $\bb R^3$ with the gauge transformation $\Phi\to-\Phi$. This is as expected, as the three-dimensional theory has to obey time-reversal invariance for all $\theta$.

While at tree level $\phi$ has a flat potential, radiatively a potential is generated~\refs{\GrossBR,\WeissRJ} with two minima at
\eqn\twovac{\beta\phi=0~,\qquad \beta\phi=\pi~, \qquad  {\rm corresponding \ to } \qquad U=\pm \unit~.}
These minima are distinct and indicate a spontaneous breaking of the 0-form center symmetry. As we decrease the temperature, the two vacua approach each other until they collide at $\beta\phi={\pi\over 2} \sim -{\pi\over 2}$.
Since $\theta$ only makes subleading contributions at very high temperatures, the conclusion that the 0-form symmetry is spontaneously broken is valid for all $\theta$.

Let us now consider the 1-form symmetry in three dimensions. The mass squared of the scalar $\phi$ in the minima~\twovac\ scales as $ \beta^{-2}$. Therefore, the scalar $\phi$ is much heavier than the dynamical scale $g_{3d}$
\eqn\mphis{m_{\phi}^2\gg g^2_{3d}~.}
We can thus safely integrate it out and the low energy theory is a pure three-dimensional $SU(2)$ Yang-Mills theory, which confines. Hence, the space-like Wilson loops in all representations on which the center $\Z_N$ acts nontrivially enjoy an area law, and the 1-form $\Z_N$ symmetry is unbroken \GaiottoKFA.
Finally, note that the CP symmetry is preserved in both vacua in~\twovac.

We can now summarize the vacuum structure in the high temperature phase and the low temperature phase
\eqn\highT{\beta\ll \Lambda^{-1} : \qquad   \matrix{  &  \theta\neq 0,\pi &\theta=0 &  \theta=\pi \cr  {\rm 1-form\ center}  & \checkmark & \checkmark &\checkmark \cr  {\rm 0-form\ center} & \times & \times & \times \cr CP & - & \checkmark&\checkmark   } }

\eqn\lowT{\beta\gg \Lambda^{-1} : \qquad   \matrix{  &  \theta\neq 0,\pi &\theta=0 &  \theta=\pi \cr  {\rm 1-form\ center}  & \checkmark & \checkmark &\checkmark \cr  {\rm 0-form\ center} & \checkmark & \checkmark & \checkmark \cr CP & - & \checkmark&\times   } }

Importantly, note that for $\theta=\pi$ the situation is consistent with the anomaly~\inflow. Indeed, at least one of the three symmetries is broken at $\theta=\pi$ both at very high and at very low temperatures. Note also that we did not find a phase where the 1-form center symmetry is broken. A priori, such a phase would be consistent with the anomaly so it is noteworthy that it does not exist in the high temperature or low temperature regimes. In section~6 we will describe a scenario where such a phase exists  at intermediate temperatures.  We remind the reader again that we assume here that at zero temperature the system continues to confine with a mass gap for all $\theta$.

\subsec{The Domain Wall at High Temperatures}

There is a two-dimensional domain wall that separates the two vacua~\twovac\ of the three-dimensional theory. We would like to determine the low-energy theory on this domain wall. Furthermore, the domain wall should be consistent with the anomaly~\inflow, and we would like to understand how this comes about.

The standard lore is that the effects of $\theta$ are vanishingly small at high temperatures. However, here the adjoint scalar needs to undergo a large field excursion of the order of the temperature in order to interpolate between the two vacua, and the theta-term~\threedtheta\ is therefore activated. This is an interesting situation where the effects of $\theta$ are not small even at high temperatures.

Far from the domain wall $U=\pm \unit$ and the microscopic $SU(2)$ symmetry is unbroken. Strong dynamics confines it and creates a mass gap.  But near the domain wall $U$ must interpolate from $+\unit$ to $-\unit$ and then the $SU(2)$ symmetry is broken to $U(1)$. Since this happens in a narrow region around the domain wall, we effectively have a $U(1)$ theory in two dimensions, which does not have physical propagating degrees of freedom.  Then, due to the term~\threedtheta\ a 2d $\theta$-term is induced.
Therefore, we can model the effect of $\theta$ on the domain wall world-volume as an Abelian 2d theta angle
\eqn\thetatwod{\theta_{2d}=\theta~.}
For $\theta_{2d}\neq \pi$ we expect the domain wall to have a unique massive vacuum and the domain wall theory to be gapped
(except for the translation zeromode, of course).

For $\theta_{2d}=\pi$ we may ask if the CP symmetry can be unbroken on the domain wall. Under the CP transformation,
the 2d theta angle will have to be shifted by $2 \pi$ and we will get a corresponding shift of the domain wall action by a counter-term
$\pi i \int B$ involving the 1-form background connection.

We thus conclude that the charge conjugation symmetry is spontaneously broken on the domain wall for $\theta_{2d}=\pi$
and that the domain wall at $\theta_{2d}=\pi$ should have a two-fold vacuum degeneracy.
Because of the counter-term shift, the kink between these two vacua carries charge $1$ under the 1-form symmetry,
much as the bulk Wilson line operator in the fundamental representation.

Hence, the domain wall at $\theta=\pi$ at high temperatures is a nontrivial TQFT consisting of two vacua interchanged by charge conjugation, and a line which interpolates between them. This system reproduces the mixed charge conjugation/1-form anomaly that the domain wall needs to carry.

This system is almost identical to the standard BF TQFT with $\bb Z_2$ symmetry, except that here orientation plays a role, suggesting that this is a nontrivial modification of the standard BF theory on unorientable spaces. This modification is described in detail in appendix A.

One can think about the domain wall theory intuitively as QED$_2$ with massive even charges and theta angle~\thetatwod. Since only even charges are present, we have a $\Z_2$ 1-form symmetry. Since the charges are massive and the gauge field has no propagating degrees of freedom, the vacuum is gapped. However, for $\theta=\pi$ we have a two-fold vacuum degeneracy~\ColemanUZ. A {\it probe} unit charge particle, i.e.\ a Wilson line, would interpolate between these two vacua. The 1-form symmetry is spontaneously broken (this is because inside the Wilson loop of a unit probe charge we have the other vacuum and hence no confining string) and CP is spontaneously broken as well. We see that this domain wall is quite rich. We will have more to say about it in Appendix D.

\subsec{The Confinement/Deconfinement Transition and the 3d Ising Model}

Let us first focus on the theory at $\theta=0$. We see from~\highT\ and \lowT\ that the order parameter for the confinement-deconfinement transition is a real field (i.e.\ the wrapped $(1,0)$ loop) which undergoes a spontaneous breaking of a 0-form $\bb Z_2$ center symmetry. It is known that the confinement-deconfinement transition is a 2nd order transition in this case. It is therefore in the same universality class as the 3d Ising model~\refs{\SvetitskyGS,\SvetitskyYE}.  We denote the temperature of the transition by $\beta_c$. The disordered phase of the Ising model corresponds to the confining (low temperature) phase in $SU(2)$ gauge theory.

The first entry in the dictionary between the Ising model and the gauge theory is that the Polyakov loop becomes the Ising spin field at long distances~\refs{\SvetitskyGS,\SvetitskyYE}
\eqn\dict{\Tr\ U=\Tr P e^{i \int_{S^1} A} \longleftrightarrow 2\sigma~.}

The space-like $(0,1)$ loop lies at the boundary of a topological $\bb Z_2$ surface operator. This surface operator implements the global $\bb Z_2$ symmetry transformation of the 3d Ising model. Its boundary has a natural interpretation also in terms of the Ising model. It is the disorder line defect
\eqn\dicti{{\rm Space-like} \ (0,1)\ {\rm loop} \longleftrightarrow {\rm monodromy \ defect}~.}
The disorder line defect in the 3d Ising model flows to a nontrivial conformal line operator at long distances~\refs{\BilloJDA,\GaiottoNVA}. Hence, the $(0,1)$  loop flows to a nontrivial conformal line operator at the confinement-deconfinement transition.

The space-like $(1,0)$ line is confined and does not appear in the Ising description. The wrapped $(0,1)$ operator is attached to a topological line operator. This line operator generates the 1-form center symmetry under which the $(1,0)$ space-like line is charged. This 1-form symmetry also does not appear in the 3d Ising model. There is no contradiction here because the 1-form center symmetry is preserved on both sides of the transition.

In the broken phase of the Ising model, there is a $\Z_2$ domain wall, which interpolates between the $\sigma=1$ vacuum at one end of space and the $\sigma=-1$ vacuum at the other end. Such a domain wall will also occur if space-time is endowed with a non-trivial background gauge field for the $\bb Z_2$ symmetry of the Ising model. This domain wall is the same as the domain wall we discussed above.

So far we considered our system in $\R^3 \times S^1$.  We can compactify one of the directions in $\R^3$ on a circle.  Then we can impose 't Hooft twisted boundary conditions on the $T^2$ formed by the original circle and this new circle.  These twisted boundary conditions represent the domain wall we discussed above stretched along the noncompact $\R^2$.  We conclude
\eqn\relation{{\rm 't \ Hooft \ twisted \ b.c.} \longleftrightarrow  {\rm Ising\  Domain \ Wall.}  }
In other words, the 't Hooft boundary conditions on the $T^2$ at high temperatures give rise to a string with a Nambu-Goto action
\eqn\NG{\int d^2\sigma \sqrt {\det h}+\cdots~, }
with $h_{\alpha\beta}=\del_\alpha Z\del_\beta Z$ being the induced metric on the domain wall in the static gauge where only the transverse coordinate, $Z$, appears.

Finally, let us consider the confining string. It is convenient to take $Y=\R\times\R^2$, and view $\R$ as time. Then the spatial slice is $\bb R^2$$\times S^1_\beta$. We can consider the wrapped $(1,0)$ loop along the $S^1$. It creates some state in the Hilbert space
\eqn\state{\Tr_F P e^{i \int_{S^1} A}\bigl|vac\big\rangle~.}
At large $\beta$ the state that is created in this way does not mix with the vacuum because it is protected by the unbroken center symmetry. (At small $\beta$ this state is in the same superselection sector as the vacuum.) Therefore, it is interesting to ask about the spectrum of energies in the superselection sector of the state~\state. This can be probed, for instance, by studying the two-point functions of Wilson loops inserted at different times. The spectrum of energies for large $\beta$ is controlled again by the Nambu-Goto action~\NG\ plus various effective field theory corrections that were studied in great detail, for example, in~\refs{\DubovskySH\AharonyIPA-\HellermanCBA} and on the lattice in~\refs{\AthenodorouCS} (and see references therein). One of course finds that for large $\beta$, the ground state energy $E_0$ is
\eqn\longstring{E_0=t\beta+\ldots~.}
where the~$\ldots$ stand for a series expansion in $1/t\beta^2$ and $t$ is the tension of the confining string.

We can make detailed predictions also about finite $\beta$ in the vicinity of $\beta_c$, where this correlation function of wrapped $(1,0)$ loops is mapped to the two point function $\langle \sigma(x)\sigma(0)\rangle$ in the 3d Ising model.

It follows from conformal invariance at the phase transition that if we denote the coefficient of the $\epsilon$ operator (i.e.\ the energy operator of the Ising model) as $\lambda$ then the energy scale in the problem is $E\sim \lambda^{1\over 3-\Delta_\epsilon}$. On the other hand, $\lambda\sim \beta-\beta_c$. (This is essentially the exponent $\nu$.) Thus,
\eqn\domainten{E\sim (\beta-\beta_c)^{1\over 3-\Delta_\epsilon}\Lambda_{QCD}^{{4-\Delta_\epsilon\over 3-\Delta_\epsilon}}\sim
(\beta-\beta_c)^{0.63}\Lambda_{QCD}^{1.63}~.}
Combining~\longstring\ with~\domainten\ and also the fact that $E_0=0$ for $\beta<\beta_c$ (since the string can mix with the vacuum) one obtains a reasonable qualitative picture for the energy of a wrapped closed string in $SU(2)$ Yang-Mills theory as a function of its length.

\newsec{The Confinement-Deconfinement Transition for Nonzero $\theta$}

Above we saw that for $\theta=0$ the deconfinement transition is of second order. Let us now consider nonzero $\theta$. The operator $\sigma$ in the Ising model~\dict\ maps to the wrapped Wilson line. Therefore, it is odd under the center symmetry (and even under CP) and hence cannot be associated to the $\theta$ angle deformation.\foot{Introducing heavy fundamental quarks would correspond to turning on $\sigma$, i.e.\ a background magnetic field in the language of the 3d Ising model.} The other relevant deformation by the $\epsilon$ operator corresponds to $\beta-\beta_c$. We can imagine that $\beta_c=\beta_c(\theta)$ and the theta dependence of the coefficient of the energy operator simply shifts the critical temperature. Therefore, for small finite $\theta$, the confinement-deconfinement transition must also be a second order transition.

\ifig\figone{Known information about the phase diagram of $SU(2)$ gauge theory as a function of $\theta$ and temperature $\beta^{-1}$.  The smooth line is a second order transition in the Ising 3d universality class.  The zigzag line is expected to be a first order transition.  There are 2 vacua on the zigzag line and one vacuum on each side of it. Later in this paper we will discuss the region with question marks.
}
{\epsfxsize2.0in\epsfbox{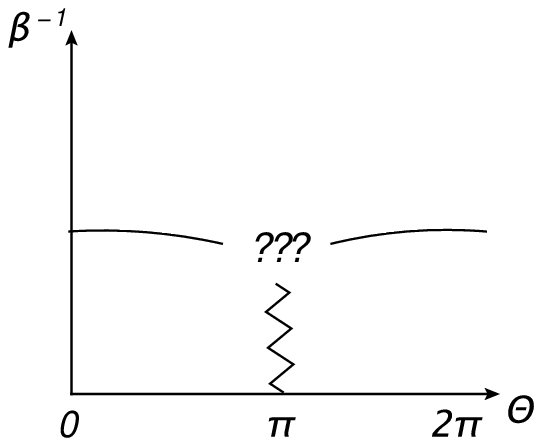}}

Given that we have already argued that CP symmetry is spontaneously broken at $\theta=\pi$ at zero temperature and given that the phase diagram has to be periodic in $\theta$, we can infer that the phase diagram is as in \figone.
The lines on the diagram have to meet in some way, which we will soon discuss.
The reflection symmetry around $\theta=\pi$ is due to CP invariance at $\theta=\pi$.

\subsec{A Mixed Gauge Theory with Dihedral Symmetry}

In order to understand better the phase diagram it is crucial to investigate the symmetries at $\theta=\pi$ further. For this, it is useful to define a ``Mixed Gauge Theory.''
This allows us to streamline and simplify the discussion considerably, but as we will explain, it is not strictly necessary to do this in order to proceed.

To define the Mixed Gauge Theory, we put the $SU(2)$ theory on a circle and promote the gauge field that couples to the 1-form $\Z_2$ center symmetry on $\bb R^3$ into a dynamical variable.
This is different from the standard gauging of the center symmetry (which leads to an $SO(3)$ gauge theory) in that we only gauge the 1-form symmetry of the 3d theory, but we do not gauge the 0-form center symmetry. Therefore, the lines that attach to the wrapped $(0,1)$ and $(1,1)$ operators disappear (and the wrapped $(1,0)$ line remains a good local operator).

Hence, the wrapped $(1,0)$, $(0,1)$, $(1,1)$ operators (and in fact all the wrapped $(a,b)$ lines)  are now good local operators on $\bb R^3$. On the other hand, all space-like loops are now attached to surfaces. This is very convenient since we can discuss more uniformly the different phases of the theory in \figone.

This mixed theory has no 1-form symmetries but it has a 0-form symmetry. The gauging of the original 1-form symmetry leads to a new 0-form symmetry.\foot{This is analogous to the fact that in 2d orbifolds, we get a new ``quantum'' 0-form symmetry. The analogous statements for higher form symmetries can be found in~\refs{\KapustinGUA,\GaiottoKFA}.}
\eqn\symmetries{\matrix{ & \theta\neq 0,\pi& \theta=0&\theta=\pi\cr {\rm 0-form \  symmetry} & \bb Z_2\times \bb Z_2 & \bb Z_2\times \bb Z_2\times\bb Z_2 & \bb D_8 } }
We see that for $\theta\neq\pi$ the bonus 0-form symmetry (which replaces the original gauged 1-form symmetry) is $\bb Z_2$. At $\theta=\pi$, due to the anomaly~\sectwoa,\inflow, the bonus $\bb Z_2$ symmetry is extended in a nontrivial way\foot{An analogous phenomenon concerning the bonus symmetries in 2d orbifolds is described in the mathematical literature in~\Deepak. We thank Y.~Tachikawa for bringing this to our attention.} over $\bb Z_2\times\bb Z_2$. Let us see how this occurs. We will give one derivation below and another one in appendix B.

As before, we denote the 2-form $\Z_2$ gauge field in 3d by $B$. When gauging the 1-form $\Z_2$ center symmetry, it is promoted to a dynamical variable and it is summer over. The gauged theory has a bonus $\Z_2$ symmetry and thus can be coupled to a background 1-form $\Z_2$ gauge field $b$. If we use the lattice formulation, the coupling to $b$ is represented by a term
\eqn\bcoupling{
 i\pi \int_Y\ b\cup B ~.
 }
 Consider now a CP transformation at $\theta=\pi$. It adds to the action a term~\sectwoa. This term can be absorbed into a redefinition of $b$:
 \eqn\CPnonan{CP: \ b\longrightarrow b+A~.}
 We see that the original mixed 't Hooft anomaly turned into a nontrivial transformation law for the field $b$ under the CP transformation. Let us denote the $\bb Z_2$ generator associated to $b$ by $h$ and the $\bb Z_2$ generator associated to the 0-form center by $c$. The CP generator will be denoted $CP$. Then~\CPnonan\ and the fact that $c,h$ generate a $\bb Z_2\times\bb Z_2$ subgroup imply that
\eqn\CPcdotc{CP \cdot c \cdot  CP =c\cdot h~,\qquad CP \cdot h\cdot  CP = h ~,\qquad ch=hc~,\qquad c^2=h^2=CP^2=1~.}
We see that $CP \cdot c$ is an element of order 4. Clearly this group is $\bb D_8$, which is the symmetry group of the square.\foot{The only other non-Abelian group at order 8, the Quaternion group, has only two elements of order 2.} We can represent these generators on the wrapped $(0,1)$ and $(1,1)$ lines, which correspond to the local operators $T(x)$, $D(x)$ in $\bb R^3$ (the wrapped Wilson line is just the product of $T$ and $D$) as follows:
\eqn\reps{CP: T\leftrightarrow D~,\qquad h: T\to -T, D\to -D~,\qquad c: D\to -D~.}
At this point, we could have chosen $c$ to act as $T\to -T$ (leaving $D$ invariant). This ambiguity has to do with the option of adding a counter-term~\addct\ before gauging. It is ultimately related to whether by gauging the generator $c$ we get $SO(3)_+$ or $SO(3)_-$ Yang-Mills theory on a circle. The choice we made in~\reps\ corresponds to $SO(3)_+$.
Note that $T,D$ transform as the two sides of a square under $\bb D_8$.

By studying the mixed gauge theory we lose no information. Indeed, by turning $b$ into a dynamical field (i.e.\ orbifolding by $h$) we can go back to $SU(2)$. If, however, we decide to turn $A$ into a dynamical gauge field (i.e.\ orbifolding by the generator $c$) then we end up with $SO(3)_+$. In order to study $SO(3)_-$ we can orbifold by the generator $h\cdot c$. In particular, the lines separating the different phases of the mixed theory, $SU(2)$, and $SO(3)_\pm$ have to be the essentially the same.

Let us summarize how to recover various gauge theories on a circle by gauging various $\Z_2$ subgroups of $\bb D_8$:\eqn\gauges{\eqalign{SU(2)&: \qquad {\rm orbifold\ by}\quad \  (T,D)\to (-T,-D)~,\cr
SO(3)_+&: \qquad {\rm orbifold\ by} \quad \  (T,D)\to (T,-D)~\cr
SO(3)_-&: \qquad {\rm orbifold\ by} \quad \  (T,D)\to (-T,D)~ .\cr
}}
Gauging these subgroup breaks the global $\bb D_8$ symmetry.\foot{In general, consider a system with symmetry group $G$. Suppose a subgroup $H$ is coupled to dynamical gauge fields. The remaining global symmetry is given by $N(H)/H$, where $N(H)$ is the normalizer of $H$; i.e.\ the maximal subgroup of $G$ within which $H$ is normal. (More precisely, if $H$ is continuous, there can also be a new bonus magnetic symmetry.) We therefore need to compute the normalizers of the $\Z_2$ subgroups, which are being gauged in~\gauges. }
In the first case, the $\Z_2$ that is gauged is a normal subgroup of $\bb D_8$, and the remaining symmetry group is  thus $\bb Z_2\times \bb Z_2$. This is exactly what we expect from the $SU(2)$ gauge theory at $\theta=\pi$. In the two other cases the $\Z_2$ symmetry that is gauged is not normal, and the quotient $N(\Z_2)/\Z_2$ is isomorphic to $\bb Z_2$. This can be understood from the fact that $\theta=\pi$ is CP invariant in $SU(2)$ but not in $SO(3)_\pm$ as they are interchanged by $\theta\rightarrow\theta+2\pi$.

In addition, gauging these 0-form symmetries also introduces a new bonus $\bb Z_2$ one-form symmetry, which we indeed expect to find in the $SU(2)$ and $SO(3)_\pm$ theories.

When we move away from $\theta=\pi$ in the mixed gauge theory we no longer have a $\bb D_8$ symmetry but just a $\bb Z_2\times \bb Z_2$ symmetry~\symmetries.
Indeed, changing the $\theta$ angle corresponds to adding some operators that break the symmetry $T\leftrightarrow D$. However, the symmetry  generated by $T\rightarrow -T, D\to D $ and $T\to T, D\rightarrow -D$ remains, as needed. If we now further orbifold as in~\gauges\  then only a $\bb Z_2$ symmetry would remain. In the $SU(2)$ case, this is just the center symmetry. In the $SO(3)_\pm$ cases, this is the dimensional reduction of the magnetic 1-form symmetry of the corresponding 4d theories.

\vskip 5cm
\subsec{The Vacuum Structure of the Mixed Theory}

\ifig\figtwo{The phase diagram in \figone\ is independent of additional gauging.  Here we add the vacua in the various regions after gauging as in the mixed gauge theory.
}%
{\epsfxsize2.7in\epsfbox{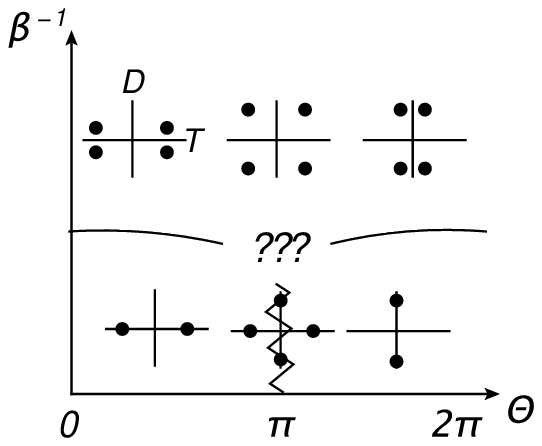}}

In \figtwo\ we present the phase diagram of the mixed gauge theory, where we also emphasize the number of vacua and their heuristic location in each of the regions. As before, we still need to determine what is happening in the region with the question marks.

The vacua are heuristically plotted on the $(T,D)$ plane and we can compute the expectation values of the Polyakov loop by taking the product $TD$.

$\theta=\pi+\delta$ and $\theta=\pi-\delta$ are related by $T\leftrightarrow D$, which is a manifestation of the Witten effect in the  mixed theory.
In the lower left region, $\bb Z_2\times\bb Z_2$ is spontaneously broken to $\bb Z_2$, generated by $D\to -D$. At low temperatures at $\theta=\pi$ each vacuum preserves either $D\to -D$ or $T\to -T$. So $\bb D_8$ is spontaneously broken to $\bb Z_2$. This includes the spontaneous breaking of the $CP$ generator. At high temperatures for $\theta<\pi$, $\bb Z_2\times \bb Z_2$ is spontaneously broken to the trivial group.  At high temperatures and $\theta=\pi$ each vacuum preserves a $\bb Z_2$. It is either the CP generator or $CP\cdot h$ in the conventions of~\reps.

We can also see that orbifolding as in the first line of~\gauges\ we are led to the correct phases of $SU(2)$ gauge theory, coinciding precisely with the phases expected in~\highT\ and~\lowT. The symmetry that we orbifold by to arrive at $SU(2)$ gauge theory is $(T,D)\to (-T,-D)$. This symmetry is always spontaneously broken in the phase diagram in \figtwo. Therefore, the resulting one-form symmetry in $SU(2)$ is always preserved in the vacuum, in agreement with~\highT\ and~\lowT.
Here we used the fact that, quite generally, if a spontaneously broken symmetry is gauged, the bonus symmetry is unbroken, and vice versa. (This is familiar from the 2d Ising model~\KadanoffKZ\ and its generalizations to higher dimensions  have been discussed, including various subtleties, in~\refs{\KapustinGUA,\GaiottoKFA}.)

\subsec{The Vacuum Structure of $SO(3)_\pm$}

\ifig\figthree{The phase diagram in \figone\ with the vacua in the $SO(3)_+$ theory in the different regions.
}%
{\epsfxsize2.7in\epsfbox{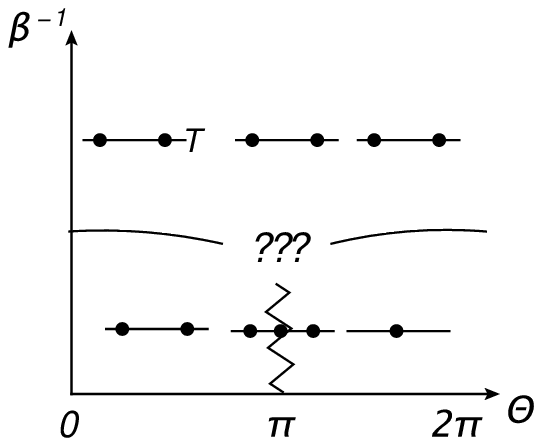}}

The phase diagram in \figtwo\ for the mixed gauge theory has immediate consequences for the phases and number of vacua in $SO(3)_\pm$ gauge theory.  Consider first the second orbifold in~\gauges, which leads to $SO(3)_+$ theory.  It is presented in \figthree.

There are several rather interesting features in this diagram. First, we plotted the vacua heuristically on the $T$ axis, which corresponds to the wrapped $(0,1)$ line,
since the wrapped lines $(1,0)$ and $(1,1)$ are projected out by the orbifold (more precisely -- they are now attached to a topological line in $Y$).

There are three vacua at $\theta=\pi$ at low temperatures. One of them preserves the 0-form center symmetry $T\to -T$. The other two vacua break this symmetry spontaneously, however, the symmetry that we gauged is unbroken in these two vacua!
This means that these vacua carry a nontrivial TQFT. Indeed, the 1-form symmetry is spontaneously broken in these vacua, which means that the space-like 't Hooft loop (which is not attached to a surface now) has a perimeter law, as we expect. This TQFT exists in the two vacua for $\theta<\pi$ while the single vacuum for $\theta>\pi$ is trivial. At high temperatures we always have two vacua, which break the 0-form center symmetry spontaneously and the 1-form symmetry is always unbroken.

For $SO(3)_-$ the picture above is reversed, with one low-temperature vacuum for $\theta<\pi$ and two for $\theta>\pi$.

This behavior of the vacua at zero temperature is consistent with the description in~\AharonyHDA.

\subsec{The CP Domain Wall}

We have seen above that there is a first order transition at $\theta=\pi$ at low temperatures, where $CP$ is spontaneously broken. Hence, there should be a domain wall at $\theta=\pi$ (at zero and also low, nonzero, temperature). Here we will discuss the properties of this domain wall.

In $SU(2)$ gauge theory the center 1-form and 0-form symmetries are unbroken at low (nonzero) temperatures at $\theta=\pi$ and hence the domain wall between the two vacua should also carry a $\bb Z_2$ 1-form and $\bb Z_2$ 0-form symmetries.
The theory cannot be trivial because of the anomaly~\inflow. Indeed, upon coupling the 0-form and 1-form symmetries to background gauge fields $A,B$, the domain wall theory should be gauge invariant only when coupling to the bulk TQFT with an action $i\pi  \int_Y A\cup B$.

A natural way to achieve this is to postulate that the domain wall theory is the $\bb Z_2$ gauge theory in two dimensions (i.e.\ $BF$ theory at level 2)
\eqn\Ztwogt{S=i\pi \int_\Sigma b\cup \delta a~.}
Here $a$ and $b$ are $\Z_2$-valued 1-cochain and 0-cochain, respectively.
This theory has two vacua distinguished by the expectation values of $e^{i \pi b}=\pm 1$.
The line $e^{i\pi \int a}$ is the domain wall between them, and its orientation is not important.

There is a quick way to see that this theory reproduces the required anomaly. We gauge the symmetry $b\to b+1$ so the two vacua become one. We are therefore allowed to consider now the vortex configuration for $b$ where $b$ returns to itself up to a shift by $1$ around some point in $\Sigma$. This operator carries charge 1 under the gauge field $a$. We can cancel the gauge charge by adding a line to it or by multiplying by some massive field in the theory. There is therefore no one-form symmetry left in the theory. The conclusion is that if we gauge the symmetry corresponding to shifts of $b$ we at the same time break the 1-form symmetry. This is the hallmark of mixed anomalies.

Another perspective on this domain wall follows from the discussion in the introduction of a domain wall between $\theta $ and $\theta +2\pi$, which has a $SU(2)_1 \sim U(1)_2$ Chern-Simons theory on it.  It is the boundary of the 4d TQFT with an action ${i\pi\over 2}\int_X B\cup B$~\GaiottoKFA.  The CP domain wall at $\theta =\pi$ is the same domain wall.  At finite temperature this bulk term becomes $i\pi \int_Y A\cup B$ and the theory on the domain wall becomes the dimensional reduction of the Chern-Simons theory; i.e.\ \Ztwogt.

We conclude that for low, nonzero temperature, the CP domain wall is the two-dimensional $BF$ theory at level 2, which can be also viewed as the $\Z_2$ gauge theory. In the deep infrared the domain wall has just two vacua and a (one dimensional) domain wall relating these two 2d vacua.

\newsec{An Inequality for the Multicritical Region}

There are two essential conceptual possibilities for the physics in the unknown region around $\theta=\pi$.
One option involves at least one point in the phase diagram where the generator $h$ in~\reps\ is unbroken. At such a point it is necessarily true that $\langle T\rangle=\langle D\rangle=0$ and hence the full $\bb D_8$ symmetry is restored. Hence, due to~\gauges\ the 1-form symmetry of $SU(2)$ is spontaneously broken and the space-like Wilson line has a perimeter law (this is again due to the fact that if an unbroken symmetry is gauged then the bonus symmetry is spontaneously broken). Therefore, the QCD string becomes tensionless. Note that in all the phases that appear in \figtwo\ there is no point with unbroken $h$. So this scenario might appear unorthodox at first sight, and indeed we do not expect this to take place for sufficiently large $N$, but it may be realized for $SU(2)$ Yang-Mills theory. We analyze such scenarios in the next sections.

In this section we  make the most conservative assumption, i.e. that there does not exist a vacuum with unbroken $\bb D_8$ symmetry. Let us study the consequences of this assumption.

Let us assume that the vertical first order line terminates (at some 2nd order transition) before it reaches the deconfinement transition line.
This is clearly inconsistent with \figtwo\ because for $\theta<\pi$, below the deconfinement transition, we have an unbroken $D\rightarrow-D$ symmetry while for $\theta>\pi$ we have an unbroken $T\rightarrow -T$ symmetry. Therefore, if the first order line terminates before it reaches the deconfinement line there must be some line where both symmetries are unbroken and hence also $h$ is unbroken, which leads to a contradiction.

We can repeat the same argument in the original $SU(2)$ gauge theory to see the connection of this argument to our anomaly. Assuming an everywhere unbroken 1-form symmetry, the anomaly implies that there cannot be a trivial vacuum with unbroken CP and 0-form center symmetry. But below the deconfinement transition the 0-form center symmetry is unbroken by definition and hence CP has to be broken.

\ifig\figfour{The phase diagram in the mixed theory (as in \figtwo), where in the region with question marks we assume that the first order line continues to higher temperatures beyond the deconfinement line.
}%
{\epsfxsize3.0in\epsfbox{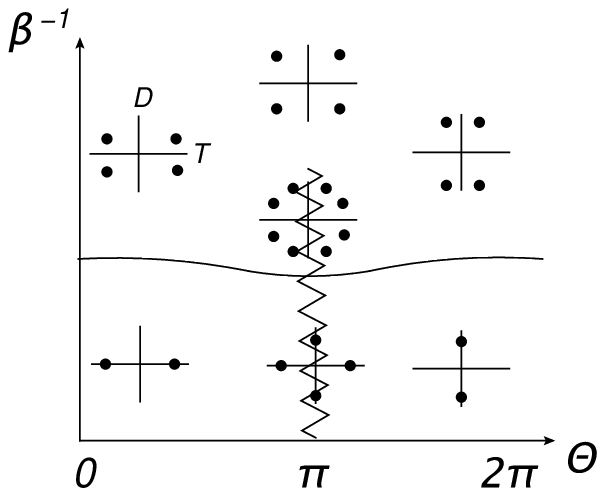}}

Therefore, we find that assuming that the space-like Wilson loop is always confined, \eqn\gen{T_{CP}\geq T_{deconf.}~,}
with $T_{CP}$ being the CP restoration temperature. As we explained,  this inequality follows from the anomaly~\sectwoa. The phase diagram (of the mixed gauge theory) would therefore look as in \figfour. It is likely that $T_{CP}=T_{deconf}$ is fine-tuned. This is especially the case in $SU(2)$ gauge theory, where the deconfinement transition is 2nd order. Below we will see that for $SU(N)$ gauge theory with $N>2$, the equality $T_{CP}=T_{deconf}$ does not require fine tuning.

Let us make some further comments about the intersection of the first-order and second-order lines. We have at that point two vacua in $SU(2)$ gauge theory and four vacua in the mixed theory. At each of these vacua, we have a three-dimensional Ising model. These vacua are separated by an energy scale of order $\Lambda_{SU(2)}$,  which is the only energy scale at this point. Therefore, there is no necessity that all the four Ising models would be simultaneously described by one three-dimensional effective field theory. The cutoff of the three-dimensional description is $\Lambda_{SU(2)}$ and so the UV completion could be just the four-dimensional theory. This is reminiscent of the two vacua in the 4d $\CN=2$ theory~\refs{\SeibergRS,\SeibergAJ}.

In the figure above the first order CP transition line ends with a second order transition in the 3d Ising universality class.

While our description here of the phases of the theory (assuming that the space-like loops are confined)  was for $SU(2)$ Yang-Mills theory, in fact, a straightforward adaptation of this scenario makes perfect sense for $SU(N)$ Yang-Mills theory. The inequality~\gen\ continues to hold for all $N$.  One important fact is that the deconfinement transition for $SU(N)$ with $N>2$ is believed to be a first order transition. In that case, the equality $T_{CP}=T_{deconf}$ does not require fine tuning.\foot{We thank J.~Maldacena for a related discussion.} Indeed, the deconfinement transition changes the physics in a discontinuous fashion and it is perfectly plausible that in this new phase the CP symmetry is never spontaneously broken.
 In addition, if the equality is satisfied, the Ising transition at the end of the spontaneous CP breaking line is no longer necessary.

  Let us now argue that this scenario requires no fine tuning. In the $T-D$ plane we are allowed to write any potential that is $\bb D_8$ invariant ($\bb D_8$ is generated by a rotation by 90 degrees and by a reflection). The general such potential is
\eqn\potential{V=\sum_k c_k \cos(4k\varphi )~.}
with $\varphi $ the angle between $T$ and $D$.  If only the term $\cos(4\varphi )$ is present, then there are four vacua, but in general, there could be 4 or 8 vacua depending on the relative coefficients of the various terms. It takes fine tuning to have more than 8 vacua. Indeed, if there are four vacua, then $\bb D_8$ is broken to some $\bb Z_2$ subgroup. Depending on the situation, it could be either a $\bb Z_2$ reflecting around an axis or a $\bb Z_2$ reflecting around a diagonal. These two scenarios are realized at low and high temperatures, respectively.  If there are 8 vacua, then $\bb D_8$ is broken completely. These are the only scenarios that do not require fine tuning.

Considering $\theta=\pi$ we can thus view the phase diagram of figure 4 as describing a transition between 4 vacua with an unbroken $\Z_2$ in the conjugacy class of a reflection around one of the axes and 4 vacua with an unbroken $\Z_2$ in the conjugacy class of a reflection around a diagonal. The transition is through a phase with 8 vacua and a completely broken $\bb D_8$.

\subsec{Domain Walls}

The domain walls at $\theta=\pi$ above the CP restoration temperature and below the deconfinement transition are  the same as described in subsections 3.3 and 4.4, respectively. The domain walls in the region above the deconfinement transition and below the CP restoration temperature could have some interesting dynamics.

It is useful to look first at the mixed gauge theory. Then $\bb D_8$ is completely broken and we have eight massive vacua.
We can pick any of the vacua, call it $v_1$, and label all other vacua $v_g$ by the non-trivial $\bb D_8$ group element $g$ mapping $v_1$ into them.

The domain walls between these vacua will generically also have a single gapped vacuum. Notice that not all domain walls
are guaranteed to exist: in principle one may interpolate between two vacua by a sequence of stable domain walls
involving other vacua of the theory.

Although the vacua are locally equivalent, pairs of vacua are not all equivalent. The seven potential domain walls between $v_1$
 and the other vacua are all physically distinct and can potentially have distinct physical properties, such as
tension. We can label the domain walls wall between vacua $v_1$ and $v_g$ as $d_g$.

Because of the broken $\bb D_8$ symmetry, the domain wall between vacua $v_g'$ and $v_{g' g}$ will have the same properties as $d_g$.
A 3d CPT transformation will both exchange the two vacua and act as a reflection in the domain wall world-volume.
Thus $d_g$ and $d_{g^{-1}}$ domain walls have identical properties up to a space reflection.
In the case at hand, the reflection relates the two domain walls labelled by the
order 4 elements of $\bb D_8$. The other group elements are of order two and the reflection
symmetry does not add more information.

Upon gauging some $\bb Z_2$ subgroup of $\bb D_8$, vacua related by gauge transformations become equivalent. Lets denote the $\bb Z_2$
generator as $x$. The domain wall $d_x$ which interpolated between $v_1$ and $v_x$ will become a dynamical string in the gauged theory.

Domain walls $d_g$ and $d_{x g}$ now interpolate between the same pair of vacua in the gauged theory.
These two domain walls are related by a space reflection if $g^2 = x$.

\newsec{A Higgs Phase at Intermediate Temperatures}

Above we saw that assuming there are no tensionless color flux tubes leads to a prediction of an inequality,~\gen. But it is worthwhile to explore the possibility that a point with unbroken $\bb D_8$ symmetry exists.

At $\theta=\pi$, the  general effective Lagrangian at distances much larger compared to the inverse temperature has to respect the $\bb D_8$ symmetry. We know that at sufficiently low temperatures the symmetry is broken spontaneously to a $\bb Z_2$ subgroup while at sufficiently high temperatures it is broken to a different $\bb Z_2$ subgroup. In between, whether or not $\bb D_8$ is restored depends on the coefficients of various terms in the effective Lagrangian~\potential. It requires no fine tuning for the region in between low and high temperatures to have an unbroken $\bb D_8$ symmetry. But if this happens, it does require the existence of new critical points on the phase diagram. We will see one concrete realization below.

There are two distinct scenarios:
\item{1.} There is some finite collection of points with unbroken $\bb D_8$.
\item{2.} There is a whole region with unbroken $\bb D_8$.

\medskip

Note that straightforward ideas like joining the three lines of \figtwo\ and having and $O(2)$ model at the intersection (with unbroken $\bb D_8$ symmetry embedded inside the $O(2)$) are inconsistent because the $O(2)$ model admits an $O(2)$ invariant deformation in which the $O(2)$ symmetry is unbroken and the vacuum is unique and trivial. This is the usual disordered phase of the $O(2)$ model. Such a phase does not exist in \figtwo.

A variation of this scenario is to postulate the existence of some 3d CFT which contains a $\bb D_8$ subgroup in its symmetry group, has two relevant deformations, but does not have a disordered phase.
Such a 3d CFT would typically be expected to have a discrete $\bb D_8$ 't Hooft anomaly (which would explain why a trivial disordered phase does not exist). But we show in Appendix C  that such an anomaly does not exist in the mixed theory.

In view of these considerations, we regard the first scenario as unlikely and turn to the second one.

\subsec{$O(2)$ Phase Diagrams}
To prepare for the construction of the phase with unbroken $\bb D_8$, let us analyze the phases of the Ginzburg –Landau theory with two fields,  $\vec M=(M_1,M_2)$
and potential
\eqn\LGH{r \vec M^2+ k (M_1^2-M_2^2)+\vec M^4+\nu (M_1^4+M_2^4)~.}
At the $O(2)$ fixed point the term $\nu$ is irrelevant and hence we hold it fixed. Our phase diagram is two-dimensional, containing $r,k$. The phase diagram crucially depends on the sign of the dangerously irrelevant parameter $\nu$. (The effects of $O(2)$ anomalous dimensions, which we essentially neglect here, are to curve the various lines in the figures below.)

\ifig\figfive{The phase diagram of the potential \LGH\ for $\nu<0$.  The zigzag line corresponds to a first-order transition
}%
{\epsfxsize3.0in\epsfbox{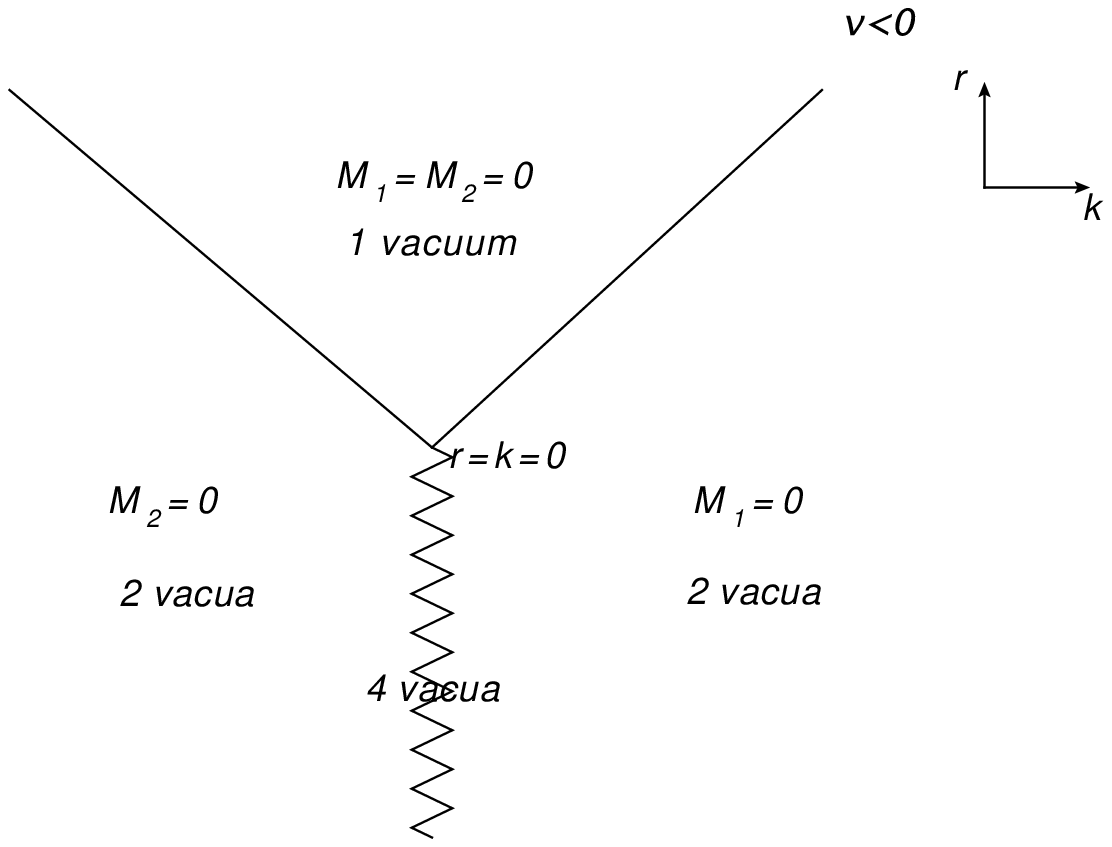}}
\ifig\figsix{ The phase diagram of the potential \LGH\ for $\nu<0$.
}%
{\epsfxsize3.0in\epsfbox{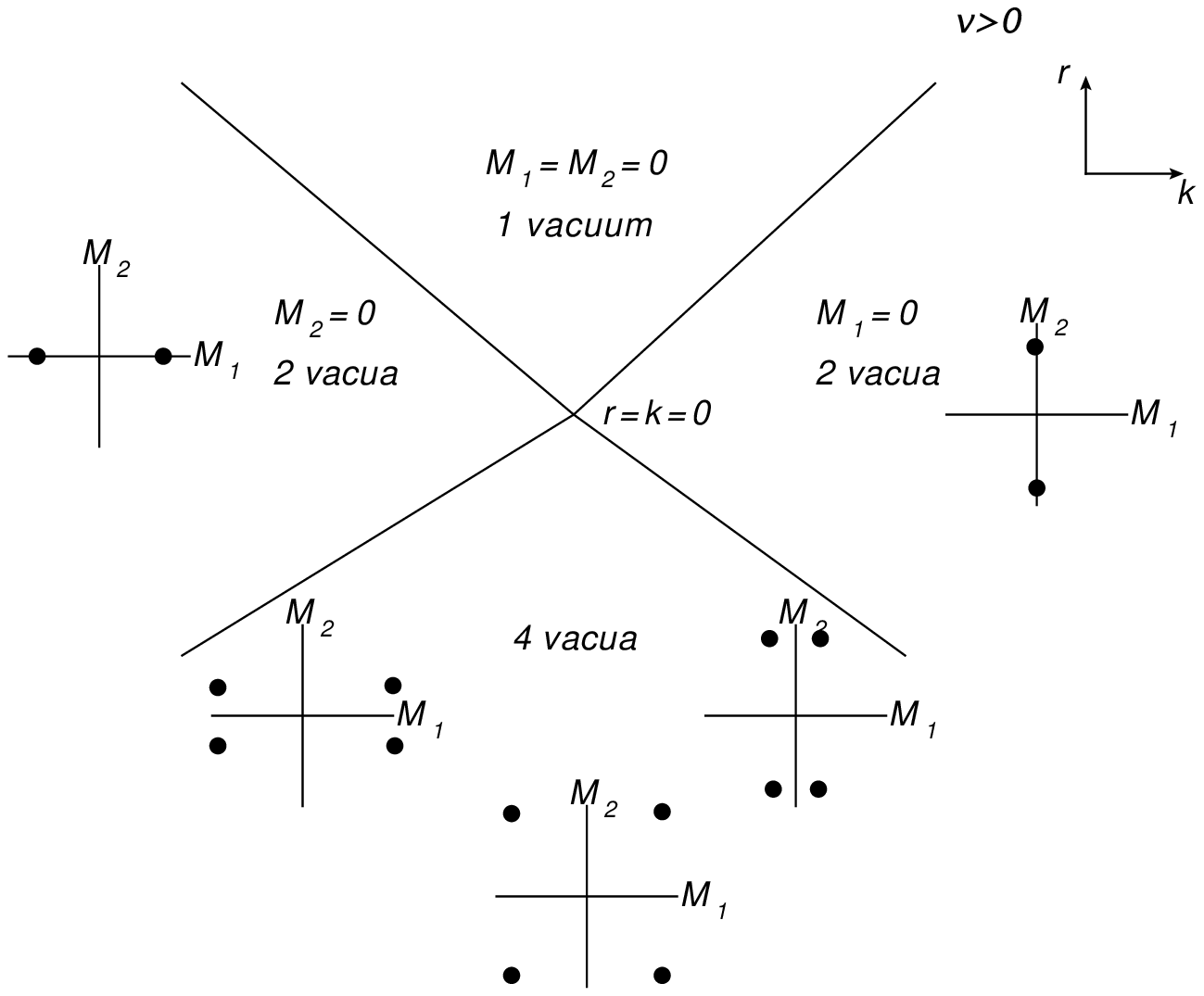}}

It is immediately clear that the phase diagram is invariant under $k\rightarrow -k$. This just exchanges $M_1$ and $M_2$, which corresponds to the $CP$ generator in~\reps. Clearly, for large positive $r>|k|$ we always get one disordered vacuum. If $r<|k|$ but $r$ is not too small (in a sense that we will specify momentarily), then we get into an ordered phases. One finds the phase diagram in \figfive\ and \figsix\ corresponding to $\nu<0$ and $\nu>0$ respectively.  The main difference between them is that the first-order line in \figfive\ splits in \figsix\ into two second-order transitions due to the dangerously irrelevant operator $\nu$.

\subsec{A Trivial Phase with Unbroken $\bb D_8$}

We have already noted above that the low temperature phase of the mixed gauge theory, \figtwo, is very similar to the bottom half of \figfive. Similarly, the high temperature part of \figtwo\ is extremely similar to the lower half of \figsix.

\ifig\figseven{A suggested phase diagram with an unbroken $\bb D_8$ phase in the shaded region.  In terms of the $SU(2)$ gauge theory, this is a Higgs phase.
}%
{\epsfxsize2.7in\epsfbox{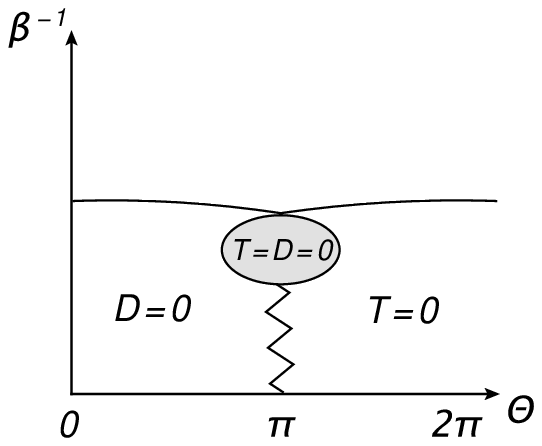}}

It is therefore suggestive to glue these two descriptions together, as in \figseven. We obtain a phase diagram with a tricritical $O(2)$ point and a tetracritical $O(2)$ point. As we explained above, both of these are locally possible multicritical points for the $O(2)$ CFT, depending on some dangerously irrelevant operators. The disordered phase is present in a compact region, where $T=D=0$, and so the fact that it is not observed at the weakly-coupled or lattice-accessible regions of the theory is not a problem.

The shaded region in \figseven\ corresponds to a single, trivial (gapped), $\bb D_8$-preserving vacuum. This is possible only if there is no 't Hooft anomaly for the $\bb D_8$ symmetry, a fact that we verify in Appendix C.
This phase can be interpreted as a Higgs phase of the $SU(2)$ gauge theory. Indeed, an unbroken bonus $\Z_2$ symmetry of the mixed theory means that the 1-form $\Z_2$ symmetry of the $SU(2)$ gauge theory is broken, and hence the spatial Wilson loops have a perimeter law. This is a hallmark of the Higgs phase. The 0-form center $\Z_2$ symmetry is unbroken, so the Polyakov loop still has zero expectation value, as do the 3d local operators $T$ and $D$.

\ifig\figeight{The phase diagram of \LGH.  Changing the temperature $\beta ^{-1}$ with fixed $\theta=\pi$ in the gauge theory is a line in this diagram starting at the lower left side and ending in the lower right side.  The dashed line corresponds to the scenario in section 5 with \figfour, and the wiggly line corresponds to the scenario in section 6 with \figseven.
}%
{\epsfxsize2.7in\epsfbox{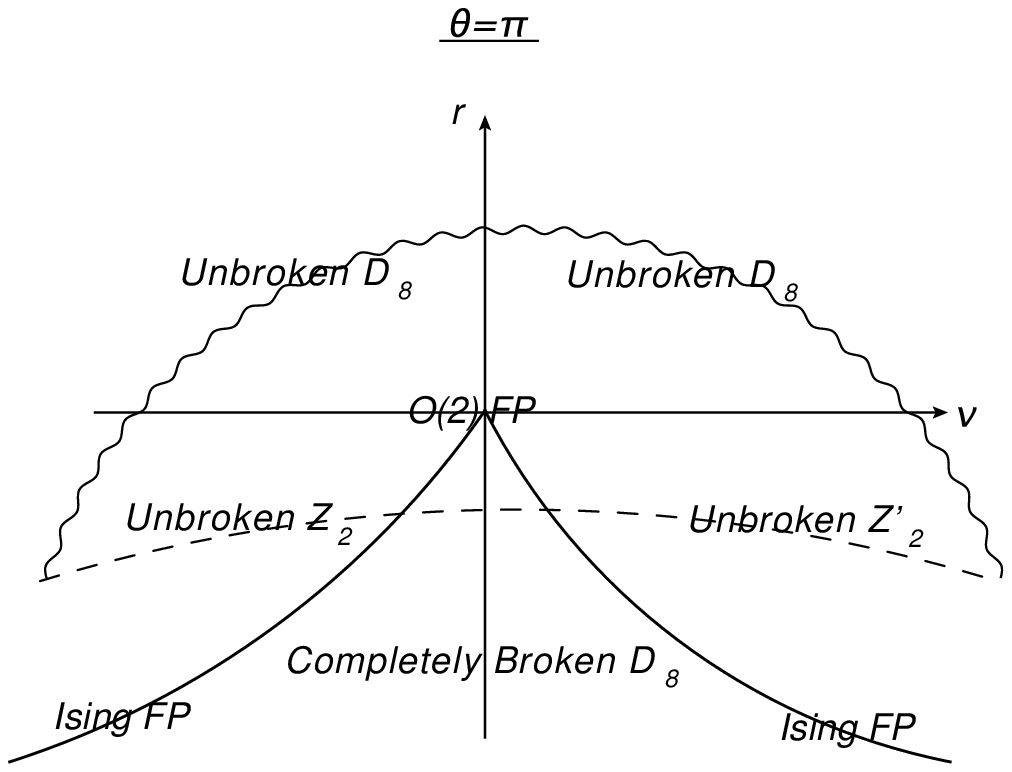}}

Whether the present scenario or the scenario of the previous section is realized depends on whether the $\bb {D}_8$ symmetry is unbroken somewhere. None of these scenarios requires fine tuning.  In order to compare them, \figeight\ depicts a slice at $\theta=\pi$. As we change the temperature, the parameters $r,\nu$ in~\LGH\ change continuously. The system has $\bb D_8$ symmetry throughout. For sufficiently negative $\nu$ and fixed $r$ we would always have four vacua and an unbroken $\Z_2$ symmetry in the conjugacy class of a reflection around one of the axes. For sufficiently large and positive $\nu$ the vacuum is again four-fold degenerate but preserves $\Z'_2$ in the conjugacy class of a reflection around a diagonal. In between, for negative $r$, there is a phase with 8 vacua and completely broken $\bb D_8$ symmetry. For positive $r$ the system always has one vacuum and unbroken $\bb D_8$ symmetry. At $r=0$ we encounter the $O(2)$ Wilson-Fisher fixed point (this is true for all $\nu$ since the corresponding operator is irrelevant in the $O(2)$ fixed point). There are also two lines with $r<0$ which support the 3d Ising model. We see that the present scenario corresponds to the wiggly curve (the temperature increases from left to right) -- it intersects the $O(2)$ model lines twice and it includes a phase with unbroken $\bb D_8$ symmetry.
The scenario of section 5 corresponds to the dashed line -- it intersects the Ising lines twice and it does not  include a phase with unbroken $\bb D_8$ symmetry.

\newsec{Scenarios with deconfinement at zero temprerature}

The scenarios outlined above assumed that $SU(2)$ Yang-Mills theory is confining at zero temperature for all $\theta$. This is suggested by the known behavior of the softly broken ${\cal N}=1$ supersymmetric Yang-Mills theory and (at large $N$) the behavior of holographic models. However, none of these arguments are conclusive, and it is conceivable, especially for $SU(2)$ Yang-Mills theory, that there exists a zero-temperature phase with a Coulomb or Higgs behavior of the Wilson loops either at $\theta=\pi$ or in some interval  $\theta\in (\pi-x,\pi+x)$ for some $0<x<\pi$. Let us discuss the possibilities for the phase diagram in such cases.

\bigskip
\centerline{\it Coulomb behavior}
\bigskip

\ifig\fignine{A possible phase diagram for $SU(2)$ gauge theories.  The second order deconfinement line touches the zero temperature line at $\theta =\pi \pm x$.   The low energy 4d theory at zero temperature with $\theta\in (\pi-x,\pi+x)$ consists of a free photon.  In the special case of $x=0$ there is a single transition at zero temperature.
}%
{\epsfxsize2.7in\epsfbox{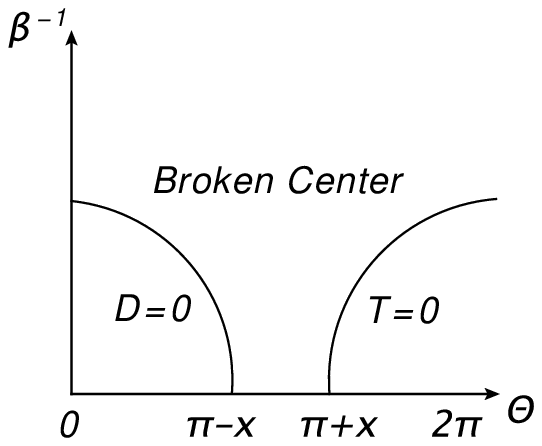}}

The simplest possibility, depicted in \fignine, is that the  deconfinement temperature decreases as one increases $\theta$, and goes to zero either at $\theta=\pi$ or at two points $\theta=\pi \pm x$ with $0<x<\pi$. In the first case the 4d theory is not confining only at $\theta=\pi$ and in the second case it is not confining in the range $\theta \in (\pi -x,\pi+x)$.\foot{For $SU(N)$ gauge theory with $N>2$ the deconfinement line is a first order line, and therefore the zero-temperature phase transition can be either first order or second order. For $SU(2)$ gauge theory, these transition points must be second order.}

If the deconfinement lines touch the zero temperature line at $\theta =\pi \pm x$, there is a critical point at these values of $\theta$.  It describe a continuous transition between a confining phase and a non-confining phase for $\theta \in (\pi-x,\pi+x)$.  The latter can be Higgs or Coulomb both with broken 1-form $\Z_2$ symmetry. Let us consider  the possibility that the physics in the interval $\theta \in (\pi-x,\pi+x)$ is in a Coulomb phase.\foot{We do not specify the CFTs at the transitions. Note that the simplest option, namely scalar QED with the scalar being a monopole or a dyon of the microscopic degrees of freedom, is ruled out because this would be a first-order transition due to the Coleman-Weinberg effect~\CWpot.}

Assuming $x\neq 0$, let us elaborate on this Coulomb phase theory.
The free photon theory has continuous magnetic and electric one-form symmetries. We imagine that the electric one-form symmetry is accidental and it is in fact entirely broken by some heavy charged particles. The magnetic one-form symmetry is broken to $\Z_2$, which matches the $\Z_2$ one-form electric symmetry of the original $SU(2)$ degrees of freedom. This symmetry acts on 't Hooft loops of the $U(1)$ gauge theory and it is spontaneously broken in the Coulomb phase.

Now let us examine the physics of this (free) theory at finite temperature.  To first approximation the low energy 3d theory consists of a free photon and a real neutral (periodic) scalar arising from the gauge field holonomy.  However, we need to add to this theory a monopole operator to reflect the fact that the 4d 1-form symmetry is $\Z_2$ rather than $U(1)$.\foot{An analogous situation was encountered in~\AharonyDHA.  There the proper reduction of a 4d theory to 3d needed the addition of a monopole operator in the 3d effective Lagrangian.} In addition, since the electric one-form symmetry is completely broken, this shift symmetry of the holonomy scalar is broken and we expect it to have a potential and be massive.

As in the famous Polyakov mechanism, the photon acquires a mass due to the monopole operator. Since the monopole operator is $\Z_2$ invariant, there are two such gapped vacua and the $\Z_2$ 0-form symmetry is spontaneously broken. The 4d 1-form symmetry (which is spontaneously broken at zero temperature) therefore leads to a spontaneously broken $\Z_2$ 0-form global symmetry and an unbroken $\Z_2$ 1-form symmetry in three dimensions.  Let us elaborate on the latter point. The $\Z_2$ 1-form symmetry in four dimensions splits to a 0-form symmetry and a 1-form symmetry.  The former corresponds to the topological $\Z_2\subset U(1)$, which is preserved by the explicit monopole operator in the low energy theory. The 1-form symmetry acts on lines in the 3d theory.  Confinement in the 3d theory means that these lines have an area law and this symmetry is unbroken.

This is precisely what we require from the finite temperature phase between $\theta=\pi\pm x$, since it is in the same universality class as the high temperature phase -- see Fig 9. In this example we see that a compactification on a very large circle can lead to a rather dramatic change in the pattern of symmetry breaking and in the spectrum. This is of course only possible if there are massless modes to start with.

It could happen that the two deconfinement transition lines touch the zero temperature line at $\theta =\pi$. This can be thought of as the $x\to 0$ limit of the previous picture and there would be no Coulomb phase in this case. This picture is reminiscent of the $O(3)$ model in two dimensions, which has a second order phase transition at $\theta=\pi$ at zero temperature (see~\refs{\AffleckCH,\Shankar} and references therein).

\bigskip\bigskip
\centerline{\it No Higgs phase at zero temperature}
\bigskip

We would like to close by making some comments about the possibility that these deconfinement lines do not simply curve down, but there is a more complicated pattern.  For example, we can have a phase diagram like \figseven\ with the Higgs region extended all the way to zero temperature.  It can touch the zero temperature line at two points $\theta =\pi \pm x$ or at a single point, $\theta =\pi$. The former scenario can actually be ruled out. Since the 0-form center symmetry is unrbroken inside the blob, it would be also unbroken in four dimensions and hence such a scenario contradicts the fact that the anomaly forces either the center symmetry or CP to be broken at $\theta=\pi$ at zero temperature (if the system is gapped). Similarly, we can exclude in this way the scenario where the deconfinement temperature vanishes at $\theta=\pi\pm x$, and there is a zero-temperature Higgs phase in the interval $\theta\in (\pi-x,\pi+x)$, with a mass gap and a spontaneously broken $\Z_2$ one-form symmetry.

\bigskip
\noindent{\bf Acknowledgments}

We would like to thank O.~Aharony, F.~Benini, C.~Cordova, M.~Dine, J.~Gomis, M.B.~Green, T.~Johnson-Freyd, M.~Metlitski, A.~Schwimmer, S.~Shenker,  and E.~Witten for useful discussions, and especially Y.~Tachikawa for collaboration at the early stage of this work.
The work of D.G. was supported by the Perimeter Institute for Theoretical Physics. Research at the Perimeter Institute
is supported by the Government of Canada through Industry Canada and by the Province of
Ontario through the Ministry of Economic Development and Innovation.
A.K. is supported by the Simons Investigator Award and in part by the U.S. Department of
Energy, Office of Science, Office of High Energy Physics, under Award Number
DE-SC0011632
Z.K.
is supported in part by an Israel Science Foundation center for excellence grant and by
the I-CORE program of the Planning and Budgeting Committee and the Israel Science
Foundation (grant number 1937/12). Z.K.\ is also supported by the ERC STG grant 335182
and by the United States-Israel BSF grant 2010/629.
NS was supported in part by DOE grant DE-SC0009988.  NS thanks the Hanna Visiting Professor Program and the Stanford Institute for Theoretical Physics for support and hospitality during the completion of this work.

\appendix{A}{A Modification of the $\Z_2$ Gauge Theory}

The standard $BF$ theory in two dimensions with $\bb Z_2$ symmetry has the following continuum action:
\eqn\Ztwogti{S={1\over \pi}\int_\Sigma b\wedge da~.}
Here $a$ is a $U(1)$ gauge field, $b$ is a $2\pi$-periodic scalar, and $\Sigma$ is an orientable 2d manifold.
This theory has a 0-form $\bb Z_2$ symmetry shifting $b$ by $\pi$,  and a 1-form $\bb Z_2$ symmetry $a\to a+\lambda$ where $\lambda$ is a flat $U(1)$ gauge field, whose holonomies are $\pm 1$.
This theory has two vacua distinguished by the expectation values of $e^{ib}=\pm 1$.
The line $e^{i\int a}$ is the domain wall between these vacua.

This theory can be extended to unorientable space-times in two different ways. One of them is obvious and the other is reminiscent of Dijkgraaf-Witten-type gauge theories~\DijkgraafPZ . It is easiest to describe it on the lattice. Then instead of continuum fields $b$ and $a$ we use cochains $b\in C^0(\Sigma,\Z_2)$ and $a\in C^1(\Sigma,\Z_2)$.
The action for this theory is
\eqn\Ztwogtii{S=i\pi  \int_\Sigma \left(b\cup \delta a+a\cup a\right)~.}
The key point is that the cup product is not supercommutative on the cochain level, so $a\cup a$ is a nonzero 2-cochain, in general.

To see that on orientable manifolds this theory is equivalent to the \Ztwogt , but not in general, let us integrate over $b$, so that $a$ is constrained to be closed modulo $2$. For such an $a$, $a\cup a$ is exact on orientable 2d manifolds, but not necessarily on unorientable ones. For example, for $\Sigma=\bb RP^2$ there is a nontrivial $\bb Z_2$ gauge field $a\in Z^1(\Sigma,\Z_2)$, and its square integrates to 1. In general, $a\cup a$ is cohomologous to $w_1\cup a$, where $w_1\in Z^1(\Sigma,\Z_2)$ is a representative of the 1st Stiefel-Whitney class $[w_1]\in H^1(\Sigma,\Z_2)$~\Hatcher . Since $[w_1]=0$ if and only if $\Sigma$ is orientable, we conclude that the theory \Ztwogtii\ is equivalent to \Ztwogti\ if and only if $\Sigma$ is orientable.

Let us discuss some properties of the theory \Ztwogtii. It is convenient to reinstate $b$ and write an action as follows:
\eqn\Ztwogtiii{S=i \pi \int_\Sigma (b\cup \delta a+w_1\cup a)~.}
This is not the same $b$ as in \Ztwogtii . In particular, if we change a representative $w_1$ by an exact cocycle, $w_1\mapsto w_1+\delta f$, $b$ transforms as $b\to b+f$. This means that $b$ takes values in the orientation bundle of $\Sigma$. So if vacua are labeled by the values of $\exp(i\pi b)$, CP exchanges the two vacua. This is precisely what happens on the center-symmetry domain wall discussed in section 3.3.

Note also that the global 1-form $\Z_2$ symmetry $a\to a+\lambda$ is broken if $\Sigma$ is unorientable. This is in fact required by the anomaly of the Yang-Mills theory~\inflow\ if we interpret the 2d TQFT \Ztwogtii\ as describing the physics of the center-symmetry domain wall. Indeed, on the domain wall the center $\Z_2$ is restored. Placing the domain wall on an unorientable background means gauging the CP symmetry, and the anomaly then requires the 1-form symmetry to be broken. On the other hand, if $\Sigma$ is orientable, the 1-form symmetry is preserved, and we can gauge it. In the gauged theory the Wilson loop for $a$ is not an observable since it is not invariant under the 1-form symmetry. The operator $\exp(i\pi b)$ becomes trivial since it is the generator of the 1-form symmetry. Thus the gauged theory has a unique vacuum and no nontrivial observables, i.e.\ it is trivial. This is consistent with the fact that in the mixed theory the center symmetry domain wall does not carry any topological degrees of freedom.

\appendix{B}{Another Derivation of $\bb D_8$ symmetry}

For the discussion in this appendix we need to introduce a gauge field for $CP$
symmetry in 3d. Indeed, from the 3d viewpoint it is an ordinary global symmetry, and hence we can couple it to a standard one-form $\bb Z_2$-valued gauge field $a$ on $Y$. The anomaly inflow was given in~\inflow. Now we define our mixed gauge theory by turning the 2-form gauge field $B$ into a dynamical field and coupling to it the 1-form $\Z_2$ gauge field $b$.

The coupling $i \pi \int_Y \ b\cup B$ is invariant under $B\to B+\delta\chi$ if $\delta b=0$, but the rest of the the theory has an anomaly under the 1-form symmetry transformation, and the constraint on $b$ should be chosen to cancel it.  Recall that the anomaly is described by the 4d action~\inflow, where $M$ is a compact oriented 4-manifold such that $\partial M=Y$. When performing a 1-form symmetry transformation as above, this 4d action varies by a boundary term
\eqn\boundaryvariation{
i\pi \int_Y a\cup A\cup\chi
}
To cancel the anomaly we must choose
\eqn\cobo{\delta b= a\cup A~.}
Here $a$ and $A$ are 1-cocycles with values in $\Z_2$, while $b$ is a 1-cochain with values in $\Z_2$.

We want to interpret this equation as  a 1-cocycle condition for 1-cochain with values in some group $G$. In general, when $G$ is an extension of some $G_0$ by an Abelian group $H$, the $G$ gauge field can be described by a pair $(b,\bf b)$, where $\bf b$ is a $G_0$ gauge field (i.e.\ a 1-cocycle with values in $G_0$), and $b$ is a 1-cochain with values in $H$ satisfying a constraint $\delta b=\omega(\bf b)$, where $\omega\in Z^2(G_0,H)$ is a 2-cocycle  describing the extension, and $\omega({\bf b})\in Z^2(Y,H)$ is its evaluation on ${\bf b}$. Applying this to equation \cobo , we see that $G$ must be an extension of $\Z_2^a\times \Z_2^A$ by $\Z_2^b$, and that the cohomology class describing the extension is the product of the generators of $H^1(\Z_2^a,\Z_2)$ and $H^1(\Z_2^A,\Z_2)$. This is the usual description of the $\bb D_8$ group.

\appendix{C}{The (Absence of) Anomalies for $\bb D_8$ symmetry}

A three-dimensional theory with $\bb D_8$ global symmetry could, in principle, have global 't Hooft anomalies classified by
\eqn\anomalieseight{
H^5({\bb D}_8 ,\Z)\simeq H^4({\bb D}_8, U(1))\simeq \Z_2\oplus \Z_2~.
}
In three-dimensional theories that arise from some four-dimensional theory, the anomaly could be in principle even larger because not all three-dimensional counter-terms arise from four-dimensional counter-terms. If the anomaly does not vanish, trivial phases with unbroken $\bb D_8$ are forbidden.

It is very useful to start from a seemingly different problem, concerning the 1-form and 0-form symmetries of the original compactified $SU(2)$ theory at $\theta=\pi$. There could be a discrete analog of the Green-Schwarz mechanism~\GreenSG, where the two-form background connection, $B$, for the 1-form center symmetry is not invariant under some 0-form gauge transformations and hence $B$ satisfies a modified Bianchi identity:

\eqn\postclass{\delta B=\eta_1 a^3+\eta_2 a^2 A+\eta_3 a A^2~.}
In this Appendix we suppress the cup product and write $a^2$ instead of $a\cup a$, etc.
The parameters $\eta_i$ are valued in $\bb Z_2$ and we have not included a term $A^3$ since the four-dimensional 2-form gauge field $B$ is closed if $a$ is set to zero. If the gauge field $a$ is nontrivial, although $B$ was closed in four dimensions, its three-dimensional counterpart may not be closed. However, it still has to be homogenous linear in $(A,B)$ and therefore we only have to discuss the possible modification of the Bianchi identity of the form
\eqn\postclassi{\delta B= a^2 A~.}

Let the 4d manifold be $X$, and suppose it is a circle fibration over a three-dimensional base $Y$.  We want $Y$ to be orientable, but the total space $X$ would be unorientable because we want the orientation of the circle to undergo a flip as we travel along non-contractible closed curves on $Y$.\foot{Mathematically, this means that the circle bundle is a unit circle bundle of a two-dimensional real vector bundle $E$ over $Y$, and this vector bundle is unorientable, with the 1st Stiefel-Whitney class $w_1(E)\in H^1(Y,\Z_2)$. Note that since $Y$ itself is orientable, i.e.\ $w_1(Y)=0$, and since for any 3-manifold $w_1^2(Y)=w_2(Y)$, we also have $w_2(Y)=0$. Therefore $w_2(X)=0$ as well. Thus this geometry restricts $X$ to be a $Pin_+$ manifold.} This class is precisely the 3d  $\bb Z_2$ gauge field for the CP symmetry, $a$.

To detect whether there is a modification of the Bianchi identity of the form~\postclassi,  we must ensure that $a^2$ is nonzero. (In general, $a^2$ could be trivial even if $a$ is a nontrivial element of  $H^1(Y,\Z_2)$. For example, on a torus every $\bb Z_2$ gauge field squares to zero.)
This restricts the choice of $Y$. The simplest possible choice is $\bb RP^3$.
It is orientable but it has nonzero $H^1(Y,\Z_2)=\Z_2$ and $H^2(Y,\Z_2)=\Z_2$. Moreover, the unique nontrivial $\bb Z_2$ gauge field on $\bb RP^3$ squares to the generator of $H^2(Y,\Z_2)$, which is exactly what we need for $a^2$ to be nontrivial.

Now we consider again our potential modification of the Bianchi identity~\postclassi. For a solution to exist, the cohomology class of $a^2A$ must be trivial. But on $\bb RP^3$, if $a$ and $A$ are both nontrivial, $a^2A $ is also nontrivial. So there is no solution for $B$, if $A$ is nontrivial (and for $a$ nontrivial). We are forced to set $A=0$.

This means that in fact $\delta B=0$ and hence the possible solutions for $B$ are parameterized by
$H^2({\bb RP}^3,\Z_2)=\Z_2$. Since $A=0$, the only $B$'s on the four-dimensional total space $X$ pull back from the base and hence $H^2(X, \Z_2)=\Z_2$.
(Intuitively, this means that on the original unorientable four-dimensional space $X$, the background two-form gauge field must not have components along the Kaluza-Klein circle.)

However, we will now show that this conclusion is false and the correct answer is $H^2(X,\Z_2)=\Z_2\oplus \Z_2$. The problem boils down to computing the cohomology of a concrete 4-manifold $X$.
One can either compute this cohomology directly or use the Gysin exact sequence~\Hatcher .
The latter can be used to show that for an arbitrary circle bundle over an arbitrary oriented 3-manifold $Y$, if the total space $X$ of the fibration is a $Pin_+$ 4-manifold, then
\eqn\HpX{H^p(X,\Z_2)=H^{p-1}(Y,\Z_2)+H^p(Y,\Z_2)~.}
That is, as far as cohomology is concerned, such a circle bundle $X$ (even though it is unorientable) behaves as if it were the Cartesian product of $S^1$ and $Y$. Thus, $H^2(X,\Z_2)=\Z_2\oplus \Z_2$ and the Bianchi identity is not modified.

Having shown that the three-dimensional two-form $B$ is closed (and hence a 2-cocycle), we can now proceed to the anomalies of the $\bb D_8$ symmetry in the mixed theory. We obtain the mixed theory by adding to the action a term $i\pi \int_Y  \ b B$ and integrating over $B$. We have to require that the system is invariant under gauge transformations of $B$. At $\theta=\pi$, due to the anomaly~\inflow,  this is only possible if we assume that $b$ is not a cocycle and hence the symmetry group is $\bb D_8$ rather than $\bb Z_2^3$. This is explained in detail in Appendix B. But we should also consider $\bb Z_2$ gauge transformations of $b$. Since $B$ is a 2-cocycle, the system is perfectly invariant under such gauge transformations (and the partition function is also invariant under the standard $\bb Z_2$ gauge transformations of $a$ and $A$) and hence $\bb D_8$ has no anomalies.

The lack of anomalies is crucially important for the consistency of the phase diagram of \figseven.

\subsec{An alternative proof}
There is an alternative way to prove the absence of anomalies for the $\bb D_8$ symmetry in the mixed gauge theory:
we can deform our theory to another theory in which the $\bb D_8$ symmetry is manifestly anomaly-free. As long as the $\bb D_8$
symmetry is preserved along the deformation, 't Hooft anomaly matching will imply the desired result.

A simple way to accomplish our objective is to add an adjoint Higgs scalar field to the four-dimensional $SU(2)$ gauge theory.
This preserves the 1-form center symmetry and CP invariance of the theory.

The original theory is recovered when the Higgs field is very massive. On the other hand, in a Higgs phase one obtains a $U(1)$ gauge theory. The $U(1)$ gauge theory is coupled to massive particles with
even electric charges and general magnetic charges, so that the naive $U(1) \times U(1)$ 1-form symmetries are broken to
the $\bb Z_2$ of the underlying $SU(2)$ gauge theory.

We can set our conventions so that the non-Abelian Wilson loop in the fundamental representation goes to Abelian loops of electric charge $\pm 1$
and the non-Abelian $(0,1)$ and $(1,1)$ loops go to Abelian loops of magnetic charge $\pm {1 \over 2}$
and appropriate electric charge.

For simplicity, we keep using the $SU(2)$ conventions for the $\theta$ angle, so that $\theta \to \theta + 2 \pi$
shifts the Abelian electric charges by twice the Abelian magnetic charges.
The Abelian gauge theory has CP symmetry at $\theta = 0$ and $\theta = \pi$ inherited from the $SU(2)$ gauge theory.

Upon compactification on the thermal circle, a $U(1)$ gauge theory can be dualized to a sigma model of
two scalar fields valued on a two-torus, whose modular parameter coincides with the complexified 4d gauge coupling.
The (exponentiated) scalar fields are simply the vevs of wrapped 4d line defects. We have an electric scalar field
$\phi_e$ associated to the wrapped Abelian Wilson loop and a magnetic scalar field $\phi_m$ associated to the wrapped Abelian 't Hooft loop.

We can take the scalars to have periodicity $2 \pi$ and encode $\tau$ in the metric on the torus.
In particular, the metric at $\theta = \pi$ involves orthogonal directions $\phi_e$ and $\phi_m - \phi_e$.
The CP symmetry is a reflection of the latter direction, i.e. $CP: \phi_m \to 2\phi_e - \phi_m$.

The 0-form symmetries which descend from the $U(1) \times U(1)$ 1-form symmetries of the
$U(1)$ gauge theory coincides with the translations of the torus. We can imagine that the
low energy effective action inherited from the $SU(2)$ gauge theory induces a metric and potential on the torus
which break translations to a single $\bb Z_2$ subgroup acting on the electric scalar field as $c: \phi_e \to \phi_e + \pi$.

Dually, the $\bb Z_2$ 1-form symmetry acts non-trivially on twist line defects around which $\phi_m \to \phi_m + 2 \pi$.
Gauging the $\bb Z_2$ 1-form symmetry to go to the mixed gauge theory has the effect of enlarging the size of the magnetic scalar field $\phi_m$
to have periodicity $4 \pi$. The dual 0-form symmetry is then $h: \phi_m \to \phi_m + 2 \pi$.

Notice that the $CP$, $h$ and $c$ generators satisfy the expected $\bb D_8$ relations. This is continuously connected to the
$\bb D_8$ symmetry of the original compactified, mixed $SU(2)$ gauge theory.

Furthermore, $\bb D_8$ acts in a completely geometric way on the scalar fields of the sigma model
and is thus not anomalous.

\appendix{D}{'t Hooft Anomalies in Quantum Mechanics}

't Hooft anomalies manifest themselves in quantum mechanics as a projective representation of the global symmetry group.  We start with a classical symmetry $G$.  Often the quantum system realizes a central extension of it $\hat G$.  The added central element $P$ obviously commutes with all the group elements in $\hat G$.  But the situation here is more specific than merely saying that the symmetry group is $\hat G$. Since $P$ does not exist in the classical theory, it commutes with all the operators in the theory.  It is central not only with respect to the other symmetry elements but with respect to all the operators.  However, it can act nontrivially on the states in the system.  But as a central element it must act in the same way on all the states in the system.

The purpose of this appendix is to discuss two examples demonstrating it and to present them in the spirit of the 't Hooft anomalies in the bulk of the paper.

\subsec{A particle on a circle}

We consider the Lagrangian (in Minkowski space, so that there is an $i$ in front of the action)
\eqn\Lpart{L=\half \dot q^2+{1\over 2\pi}\theta \dot q~}
with compact $q\simeq q+2\pi$.  Since for all field configurations on a Euclidean circle $\int \dot q \in 2\pi$$ \bb Z$, $\theta$ leads to the same physics as $\theta+2\pi$.

The conjugate momentum is $\Pi_q=\dot q+{1\over 2\pi}\theta$ and the Hamiltonian is just \eqn\Hpart{H=\half(\Pi_q-{1\over 2\pi}\theta)^2~.}
We can see that $H_{\theta}$ and $H_{\theta+2\pi}$ are related by a unitary operator $e^{iq}$ which shows more formally that they are equivalent.

The appropriate differential operator to diagonalize is
\eqn\Schparticle{\half\left(-i\del_q-{1\over 2\pi}\theta\right)^2\Psi_n=E_n\Psi_n~,}
hence,\foot{Note that the wave functions are periodic under $q\to q+2\pi$. We could also give them some phase under this transformation, which is equivalent to adding a background gauge field for the phase of the wave function. Equivalently, we can say that $\theta$ corresponds to some magnetic flux inside the circle. Here we choose to use the parameter $\theta$ and consider only periodic wave functions.}
\eqn\wavespec{\Psi_n=\langle q|n\rangle={1\over \sqrt{2\pi}}e^{inq }~,\qquad E_n=\half \left(  n-{\theta\over 2\pi}\right)^2~.}

We see that the spectrum is  invariant under $\theta\to\theta+2\pi$ (this of course follows from the fact that $H_\theta$ and $H_{\theta+2\pi}$ are unitarily equivalent) but this occurs in a nontrivial way (level crossing). Note that for $\theta=\pi$ and only in this case the ground state is two-fold degenerate. We will soon see that these two states are related by a discrete symmetry, i.e.\ there is a spontaneously broken discrete symmetry (even though we are talking about quantum mechanics, where, naively,  discrete symmetries cannot be broken).

Since the particle is free, the spectrum is in representations of $U(1)$, corresponding to $\CO_\alpha : \quad q\to q+\alpha$. At $\theta=\pi$ another generator exists, charge conjugation $\CO_C$, which maps $q\to -q$. This generator is a symmetry of the Lagrangian because $\theta=\pi$ and $\theta=-\pi$ are the same according to the arguments above.\foot{As in our discussion in footnote 7, we can use time reversal and charge conjugation interchangeably since their product is always a manifest symmetry.  } It is a symmetry both for $\theta=0$ and $\theta=\pi$. At $\theta=0$ the action is simply $|n\rangle\to|-n\rangle$. The ground state is unique and it corresponds to the constant wave function, invariant under all the symmetries.
This is the standard picture in QM.

For $\theta=\pi$ the story is more involved (this corresponds to half integer flux through the circle). The instantons have alternating signs and they can partially cancel each other.
One can convince oneself that the action of the $U(1)$ symmetry and of charge conjugation on the Hilbert space at $\theta=\pi$ are
\eqn\actioni{q\to q+\alpha: \ \ |n\rangle\to e^{i\alpha n}|n\rangle~,\qquad q\to -q: \ \  |n\rangle\to |-n+1\rangle~. }

We would like to compute now the transformations generated by $\CO_\alpha$ and $\CO_{C}$:
\eqn\generated{\CO_\alpha \CO_C: \ \ |n\rangle\to  e^{i\alpha(-n+1)}|-n+1\rangle ~,\qquad \CO_C \CO_\alpha: \ \ |n\rangle\to  e^{i\alpha n}|-n+1\rangle }
It is also useful to note now that
\eqn\generated{\CO_C\CO_\alpha \CO_C: \ \ |n\rangle\to e^{-i\alpha n+i\alpha}|n\rangle ~. }
The group generated by $\CO_\alpha$ and $\CO_C$ is therefore not $O(2)$ (which is what we might have expected based on classical considerations) since the above generator does not coincide with $\CO_{-\alpha}$. We therefore introduce a new generator, which is central, $I_\beta$ which acts by
\eqn\central{I_\beta: \ \  |n\rangle\to e^{i\beta}|n\rangle~.}
We call it central because it clearly commutes with $\CO_\alpha $ and $\CO_C$. If we ignore this central element and consider~\generated\ modulo a scalar action on the Hilbert space, we get a  projective representation of $O(2)$. Therefore, $O(2)$ is realized projectively.

We therefore see that
\eqn\projOtwo{\CO_C \CO_\alpha \CO_C=I_\alpha \CO_{-\alpha}~.}

We could consider the combined generator $V_{\alpha}= I_{-\alpha/2} \CO_\alpha $ with $\alpha \in [0,4\pi)$ and see that
\eqn\newrel{\CO_CV_\alpha \CO_C=V_{-\alpha}}
so now we can forget about $I_\alpha$ but we have to identify the central extension element $P$ as $V_{2\pi}=I_\pi =-1$.  We therefore get a double cover of the group $O(2)$, namely $Pin(2)$. This should be viewed as a central extension of $O(2)$ by $P$.

Central extensions are common manifestations of anomalies (this is very familiar in two dimensions, where the Kac-Moody and Virasoro algebras are centrally extended). One should therefore conclude that there is a 't Hooft anomaly in the $O(2)$ symmetry. We will see that this anomaly has many of the usual consequences of anomalies, including anomaly matching in a nontrivial ground state, inability to couple to classical gauge fields, anomaly inflow etc.

Another perspective on the system is obtained by coupling \Lpart\ to a classical background $U(1)$ gauge field $A_0$
\eqn\LpartA{L=\half (\dot q+A_0)^2+{1\over 2\pi}\theta (\dot q+A_0) + k A_0~}
where $kA_0$ can be thought of as a ``Chern-Simons'' counter term for $A_0$.  Clearly for consistency we need $k\in \Z$. In this form it is clear that for nonzero $A_0$ the system with $\theta$ is not the same as with $\theta+2\pi$.  Instead $(\theta,k) \sim (\theta+2\pi,k-1)$.  This is similar to the more complicated $4d$ situation we encounter in the rest of the paper where a shift of $\theta$ by $2\pi$ does not leave the system invariant, but changes a counter term for a background field.  In this form it is clear that for nonzero $A_0$ the $\theta =k=0$ system is charge conjugation invariant, but for $\theta =\pi$ this symmetry is absent.

We can compactify the Euclidean time direction and compute the partition function in the presence of a $U(1)$ chemical potential $\mu=\oint A\sim\mu+2\pi$. Then the partition function due to the vacuum Qubit is
\eqn\parfun{Z_{vac}=1+e^{i\mu}~.}
We neglect the contributions from the excited states, as they would not affect the discussion below.
Note that~\parfun\  is  not charge conjugation invariant. The $\bb Z_2$ charge conjugation symmetry is $\mu\to -\mu$. This is clearly broken by the partition function.
We can say that under a charge conjugation the partition function transforms by
\eqn\quantone{Z(-\mu)=e^{-i\mu}Z(\mu)~.}
This can be interpreted as adding a counter-term
\eqn\Counterpart{\delta S=\int A~,}
which is a correctly quantized 1d Chern-Simons term.

To make the partition function charge conjugation invariant we could attempt to add a half-integer Chern-Simons term -- $k=\half$ in \LpartA. Indeed, suppose we replaced~\parfun\ by
\eqn\replaced{Z(\mu)=e^{-i\mu/2}+e^{i\mu/2}~.}
This would be charge conjugation invariant, but inconsistent with $\mu\simeq\mu+2\pi$.
We thus clearly see that there is an $O(2)$ anomaly.

Here are some possible additional points of view on this theory.
\item{1.} We can interpret~\replaced\ as giving the vacuum half-integer charge. This would mean that the symmetry group is not $O(2)$ but its double cover, which is precisely what we found through the computation of the central extension.
\item{2.} There is a way to write a consistent partition function invariant under all the symmetries, but we have to add a two-dimensional bulk. This is similar to the mechanism of anomaly inflow (topological insulator). Indeed, the half-integer Chern-Simons term $\delta S= \half \int A$ is not a well defined object in one dimension, but we can view it as resulting from the bulk integral
    \eqn\Bulkpart{\half \int_{{\cal M}_2} F~,}
    where the two-manifold  ${\cal M}_2$ ends on our the ``time line'' of our quantum system. This depends on the choice of  ${\cal M}_2$, but once we choose such an  ${\cal M}_2$ it is well defined.

\medskip

We conclude that as long as we restrict the theory to be one-dimensional, $O(2)$ is a projectively realized symmetry and the vacuum breaks it spontaneously. The $O(2)$ is centrally extended to its double cover and if we couple the system to a $U(1)\subset O(2)$ gauge field charge conjugation is necessarily broken. This conclusion can be avoided if we couple the system to a two-dimensional bulk.

It is worth noting that the situation for a massless fermion in three dimensions is entirely analogous. We can either couple it to a $U(1)$ gauge field and thus break time reversal invariance (this is what happens when we choose the fermion path integral phase to be given by a half of the $\eta$-invariant, see~\WittenABA) or we can add a four-dimensional bulk with $\theta=\pi$ and retain the time reversal symmetry as well as the gauge  symmetry.

\bigskip
\centerline{\it Adding Interactions}

A generic potential on the circle, which is some function of $\sin(q)$ and $\cos(q)$, would explicitly break the $O(2)$ symmetry completely and there would be generically one ground state.
It is interesting to leave some subgroup of $O(2)$ unbroken. Let us consider a potential which is only a function of $\cos(2q)$,
\eqn\Vpartic{V=V(cos(2q))~.}
One can have in mind some polynomial in $\cos(2q)$. This preserves reflections and rotations by $\pi$ and so we have a symmetry $\bb Z_2\times \bb Z_2$. Such a potential generically has two or four minima, depending on the relative coefficients of the polynomial.

As proven below, the quantum theory at $\theta=\pi$ still maintains some exact two-fold degeneracy. So the symmetry is spontaneously broken to $\bb Z_2$.

In the notation above we now have $\CO_\pi$, $\CO_C$, and due to~\projOtwo\ we have to also use $I_\pi$. The group is therefore
\eqn\Deight{\CO^2_\pi=\CO_C^2=I_\pi^2=1~,\qquad \CO_C \CO_\pi \CO_C= I_\pi \CO_\pi~, }
which shows that $\CO_C\CO_\pi$ is an element of order 4, while $\CO_\pi I_\pi$ and $\CO_C I_\pi$ are elements of rank 2.  Hence we identify this with the $\bb D_8$ group (we identify this as $\bb D_8$ since it has 6 elements of rank 2).

Since the central extension is by $I_\pi=-1$, the ground state is indeed guaranteed to be a nontrivial representation proving our assertion above.\foot{Proof: since $\CO_C\CO_\pi = -\CO_\pi \CO_C$ if the ground state was a one-dimensional Hilbert space, all these elements would have to be represented by some phases and hence we would get a contradiction. } The instantons that are normally expected to remove the degeneracy and leave one symmetric vacuum cancel here because of the signs with which we multiply instanton factors.

One place where we can find this model is by thinking of $q$ as  the sphaleron degree of freedom in Yang-Mills theory, $q={1\over 4\pi}\int_{\Sigma_3} A\wedge dA+{2\over 3} A^3$. Then CP becomes $q\to -q$ and the 1-form $\Z_2$ symmetry acting on $A$ becomes $q\to q+\pi$. The identification $q\simeq q+2\pi$ is just due to the usual gauge transformations in the homotpoy classes of $S^3\to SU(N)$. Equivalently, only the exponential of the Chern-Simons term is well defined. So we can interpret our quantum mechanical model as the sphaleron theory. It preserves a $\Z_2\times \Z_2$ symmetry. The two vacua and the anomaly are indicative of the anomaly and spontaneous breaking of CP in the original theory.

For a connection between the above-discussed model with hidden supersymmetry in purely bosonic systems see~\refs{\JakubskyKI,\CorreaJE}. We now describe the connection to 1+1 dimensional $U(1)$ gauge theory.

\bigskip
\centerline{\it Connecting with $QED_2$}

Consider free Abelian gauge theory in two dimensions
\eqn\Abetwod{S=\int d^2x {-1\over 4e^2}F^2+{\theta\over 2\pi} \int  F~.}
This model has no degrees of freedom on $\bb R$ but let us study it a space manifold which is a circle of size $L$. Then we can pick the gauge $A_0=0$ and remain with
\eqn\Bulktwod{S=\int d^2x \dot A_1^2+{\theta\over 2\pi} \int  \dot A_1~.}
The residual gauge transformations $A_1\to A_1+\del_1 \Omega(x_1)$. This can be used to turn $A_1$ into a constant  mod $2\pi$ (but the constant cannot be removed). In other words $e^{i \int_{S^1}  A_1}$ is gauge invariant. In terms of (dimensionless) $q=\int A_1$, which is identified with $q+2\pi$, we therefore have
\eqn\twodbound{S={1\over e^2 L}\int dt\  \dot q^2+{\theta\over 2\pi} \int  \dot q~.}
The mass of the quantum mechanical particle is therefore $\sim e^2L$.

This is precisely the theory \Lpart\ with the spectrum \wavespec. In particular, for $\theta =\pi$ there are two degenerate ground states. They can be viewed as the configurations $E=\pm \half$ for the electric field.

Now let us add massive charged particles and suppose that the two-dimensional $QED_2$ model has a one-form $\bb Z_2$ symmetry. In other words, all the charges are even.
Then we have no dynamical charge 1 particles and the two ground states cannot mix even in finite volume. This is the twofold degeneracy found above. If we integrate out the even charge particles, we remain with a theory that has charge conjugation and $\bb Z_2$ symmetry and a potential like \Vpartic, which is some polynomial in $\cos(2q)$.

Another interesting consequence is that $QED_2$ at $\theta=\pi$ with only even charges has a mixed charge-conjugation/1-form anomaly which upon a circle compactification manifests itself as the quantum mechanical model above.

\subsec{A non-Abelian example with a discrete $\theta$-parameter}

Here we consider a simple example with an anomaly associated with a discrete $\theta$-parameter.  We study a quantum mechanical particle moving on the $SO(3) = \Sn^3/\Z_2$  group manifold.

One way to think about it is to start with $4$ scalars $q^i$ with a constraint $q^i q^i=1 $ leading to $\Sn^3\cong SU(2)$.  This system has an $O(4) \cong \Z_2\sdtimes (SU(2)_L\times SU(2)_R)/\Z_2$  global symmetry. Next, we
gauge the $\Z_2$ symmetry generated by the nontrivial central element $P\in SO(4)$; i.e.\ we orbifold by this symmetry leading to the target space $ \Sn^3/\Z_2$.  After the gauging the global symmetry is $\Z_2 \sdtimes SO(3)_L\times SO(3)_R$.  All the operators transform under this symmetry; there are no operators in $SO(4)=SU(2)_L\times SU(2)_R/\Z_2$ representations with half-integer isospin in the two $SU(2)$ factors.

The target space of the system is not simply connected and therefore there is a discrete $\theta$-parameter, which we will denote by $\theta =0 $ or $\pi$.

The Hilbert space is of the form
$\CH=\oplus_j R_j$ where $R_j$ is the $(j,j)$ representation of the $SU(2)_L\times SU(2)_R$ symmetry.
Before the $\Z_2$ gauging we had all integer and half-integer values of $j$.
After the gauging all the operators have $j_L,j_R\in \Z$ and there are no transitions between states with integer $j$ and states with half-integer $j$.  This is consistent with the fact that the Hilbert space depends on $\theta$
\eqn\thetazH{\eqalign{
&\CH_{\theta=0} = \oplus_{j\in \Z} R_j \cr
&\CH_{\theta=\pi} = \oplus_{j\in \Z+\half} R_j~.}}
The nontrivial element in the $\Z_2$ orbifold, $P$ acts as $+1$ on $\CH_0$ and as $-1$ on $\CH_\pi$.

Now we want to couple the system to background gauge fields.  In quantum mechanics we compactify Euclidean time and all the information in the background field is in the holonomy $U$ around that circle.  The effect of this holonomy is that we compute traces with an insertion of the group element $U$.  Let us start with $SO(4)$ gauge fields.  There is no problem computing such traces with any $U$ either in $\CH_0$ or in $\CH_\pi$ in \thetazH.

Next, we try to repeat this with background $SO(3)_L\times SO(3)_R$ gauge fields.  Let us compare the traces with insertions of the two different group elements $U$ and $U'=UP$ in $SO(4)$, which are identified in $SO(3)_L\times SO(3)_R$.  When working in $\CH_0$ these traces are the same.  However, in $\CH_\pi$ a trace with an insertion of $U$ differs by a sign from a trace with an insertion of $U'$.  This means that the answers with background $SO(3)_L\times SO(3)_R$ fields are gauge invariant in $\CH_0$ but are not gauge invariant in $\CH_\pi$.  $SO(3)_L\times SO(3)_R$ suffers from an 't Hooft anomaly in $\CH_\pi$.\foot{It is often stated that 't Hooft anomalies are diagnosed by making the gauge field dynamical and checking whether the Hilbert space is empty, or equivalently checking whether the partition function vanishes.  If 't Hooft anomalies exist this conclusion is right.  But the converse is not true.  It is clear in our case that gauging the $SO(4)$ symmetry leaves one invariant state in $\CH_0$ and no invariant state in $\CH_\pi$.  But evidently, there is no 't Hooft anomaly in $SO(4)$.}

As is common in situations with 't Hooft anomaly, we can fix the anomaly by extending the $SO(3)_L\times SO(3)_R$ gauge fields to a 2d manifold $\Sigma$ and adding a bulk term
\eqn\bulktermna{i\pi \int_\Sigma \Big(w_2 (SO(3)_L )+w_2 (SO(3)_R )\Big)}
with $w_2$ the second Stiefel-Whitney class.
This preserves the $\Z_2$ that exchanges them.  For a closed manifold $\Sigma$ with $SO(4)$ gauge fields the two terms in \bulktermna\ are equal (modulo $2\pi i $), and \bulktermna\ is trivial.  However, now we can allow nontrivial bundles in one of the $SO(3)$ factors and not in the other and then \bulktermna\ is nontrivial.

\listrefs

\bye